\documentclass[journal=jacsat,manuscript=article]{achemso}
\usepackage{pdfpages}

\usepackage{graphicx}
\usepackage{units}
\usepackage{dcolumn}
\usepackage{amsmath}
\usepackage{verbatim} 
\usepackage{bm} 

\makeatletter
\newcommand\footnoteref[1]{\protected@xdef\@thefnmark{\ref{#1}}\@footnotemark}
\makeatother

\newcommand{\ket}[1]{|#1 \rangle}

\newcommand{\Eq}[1]{equation~(\ref{#1})}

\newcommand{\Fig}[2]{Figure~\ref{#1}{#2}} 
\newcommand{\Figs}[3]{Figures~\ref{#1}{#2} - \ref{#1}{#3}} 
\newcommand{\fig}[1]{\ref{#1}}
\newcommand{\Panel}[1]{panel~({#1})}
\newcommand{\panel}[1]{({#1})}
\newcommand{\panellabel}[1]{({#1})}

\newcommand{\figopt}{t}

\usepackage[normalem]{ulem} 

\newcommand{\clr}{Gray}
\newcommand{\suppmat}{\textcolor{\clr}{Supporting Information}}

\newcommand{\cut}[1]{\textcolor{green}{\textbf{\sout{#1}}}}  
\newcommand{\new}[1]{\textcolor{blue}{{#1}}}
\newcommand{\New}[1]{\textcolor{blue}{{#1}}}
\newcommand{\todo}[1]{\textbf{\textcolor{green}{[{#1}]}}}
\newcommand{\todox}[1]{\textbf{\textcolor{red}{[{#1}]}}}

\renewcommand{\cut}[1]{}
\renewcommand{\clr}{black}
\renewcommand{\new}[1]{#1} \renewcommand{\New}[1]{#1}
\renewcommand{\todo}[1]{} \renewcommand{\todox}[1]{}

\title{
  Switchable coupling of vibrations to two-electron carbon-nanotube quantum dot states 
}
\author{P. Weber}
\affiliation{2$^\text{nd}$ Institute of Physics, RWTH Aachen University, 52056 Aachen,
Germany}
\alsoaffiliation{JARA\,--\,Fundamentals of Future Information Technology}
\altaffiliation{ICFO-Institut de Ciencies Fotoniques, Mediterranean Technology Park, 08860 Castelldefels, Barcelona, Spain}
\altaffiliation{Equal contribution}

\author{H. L. Calvo}
\affiliation{Institute for Theory of Statistical Physics, RWTH Aachen  University, 52074 Aachen,  Germany}
\alsoaffiliation{JARA\,--\,Fundamentals of Future Information Technology}
\altaffiliation{Instituto de F{\'i}sica Enrique Gaviola (IFEG-CONICET) and FaMAF, Universidad Nacional de C{\'o}rdoba, Ciudad Universitaria, 5000 C{\'o}rdoba, Argentina}
\altaffiliation{Equal contribution}

\author{J. Bohle}
\affiliation{Institute for Theory of Statistical Physics, RWTH Aachen University, 52074 Aachen,  Germany}
\alsoaffiliation{JARA\,--\,Fundamentals of Future Information Technology}

\author{K. Go\ss}
\affiliation{Peter Gr{\"u}nberg Institute, Forschungszentrum J{\"u}lich, 52425 J{\"u}lich,  Germany}
\alsoaffiliation{JARA\,--\,Fundamentals of Future Information Technology}
\altaffiliation{Physikalisches Institut, Universit{\"a}t Stuttgart, Pfaffenwaldring 57, Stuttgart, Germany}

\author{C. Meyer}
\affiliation{Peter Gr{\"u}nberg Institute, Forschungszentrum J{\"u}lich, 52425 J{\"u}lich,  Germany}
\alsoaffiliation{JARA\,--\,Fundamentals of Future Information Technology}

\author{M. R. Wegewijs}
\affiliation{Peter Gr{\"u}nberg Institute, Forschungszentrum J{\"u}lich, 52425 J{\"u}lich,  Germany}
\alsoaffiliation{Institute for Theory of Statistical Physics, RWTH Aachen University, 52074 Aachen,  Germany}
\alsoaffiliation{JARA\,--\,Fundamentals of Future Information Technology}

\author{C. Stampfer}
\email{stampfer@physik.rwth-aachen.de}
\affiliation{2$^\text{nd}$ Institute of Physics, RWTH Aachen University, 52056 Aachen,
Germany}
\alsoaffiliation{Peter Gr{\"u}nberg Institute, Forschungszentrum J{\"u}lich, 52425 J{\"u}lich,  Germany}
\alsoaffiliation{JARA\,--\,Fundamentals of Future Information Technology}

\begin{document}
\begin{abstract}
We report transport measurements on a quantum dot in a partly suspended carbon nanotube.
Electrostatic tuning allows us to modify and even switch ``on'' and ``off'' the coupling to the quantized stretching vibration across several charge states.
The magnetic-field dependence indicates that only the two-electron spin-triplet excited state couples to the mechanical motion,
indicating mechanical coupling to both the valley degree of freedom and the exchange interaction, in contrast to standard models.

\end{abstract}
%
%
Carbon nanotubes are found to be an ideal playground for nano-electromechanical systems (NEMS)
since their high-quality, quantum-confined electronic states
are accessible by 
 transport spectroscopic techniques
and couple strongly to the excitations of different mechanical modes.
The growing interest in NEMS is fueled by the desire to accurately sense small masses and forces~\cite{Moser2013},
 address quantum-limited mechanical motion~\cite{Teufel09},
and integrate such functionality into complex hybrid devices~\cite{Viennot14},
leading to new applications~\cite{Schneider12}.
The central question is the strength of the coupling of electronic states to the vibrational modes.
Whereas molecular junctions display such modes also in electrically gated transport measurements~\cite{Park00,Pasupathy04,Osorio07a}, carbon-nanotube (CNT) quantum dots allow for a much more viable fabrication, higher mechanical Q-factors, and better tuneability as NEMS~\cite{Sapmaz05,Steele09,Leturcq09,Pei12,Benyamini14,Moser14}.
%
%
Also, the coupling to the bending mode can be combined~\cite{Ohm12,Palyi12a} with the spin-orbit (SO) interaction~\cite{Kuemmeth08,Jespersen11} by making use of the recently demonstrated~\cite{Pei12} curvature-induced SO-coupling in CNTs.\cite{Flensberg10a}
Whereas the frequency of the vibrational modes has been demonstrated to be tuneable~\cite{Sazonova04,Eichler12,Barnard12},
 another desirable feature is the ability to switch ``on'' and ``off'' the electron-vibration coupling in the same device, e.g., in envisioned quantum-information processing schemes~\cite{Palyi12a,Palomaki13}. This is also helpful for fundamental studies of systems in which mechanical motion is combined  with other degrees of freedom, e.g, the spin~\cite{Ganzhorn13} and the valley~\cite{Palyi11a}.

Recently, switchable coupling to a classical flexural mode of a CNT has been demonstrated~\cite{Benyamini14}.
In this letter, we present a CNT quantum dot NEMS with a coupling of the electronic states to a longitudinal \emph{stretching} vibration of about 200 GHz that can be turned {``on''} and {``off''}. We illustrate the advantage of this by transport measurements in the two-electron quantum-dot regime and find that the well-known Anderson-Holstein scenario breaks down in an unexpected way: Different spin states exhibit different coupling strengths to the vibrational mode.
\begin{figure*}[\figopt]
  \includegraphics[width=1.0\textwidth]{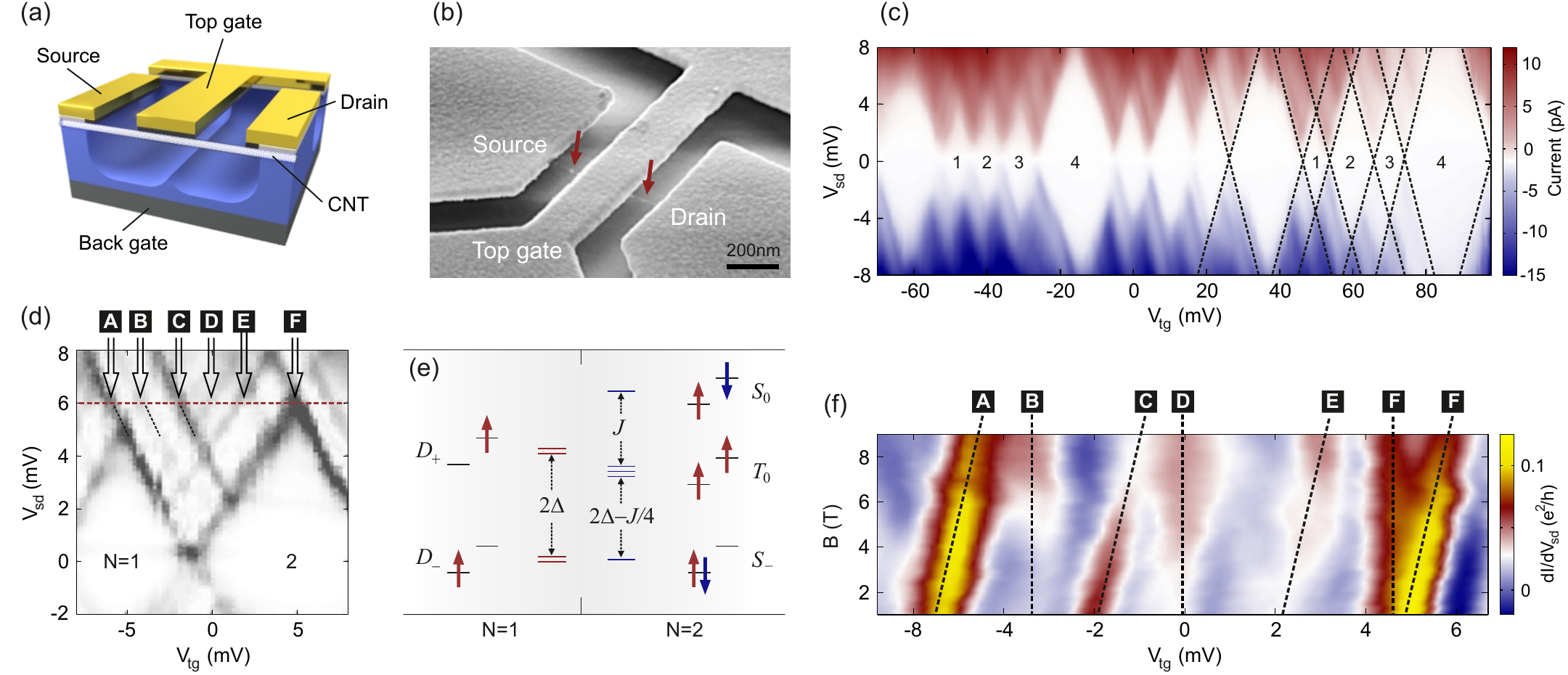}
  \caption{
    {Carbon-nanotube quantum dot characterization.}
    \panellabel{a}
    Schematic illustration of the cross-section of a partly suspended CNT connected to source and drain electrodes \new{(5nm Cr, 50 nm Au).}
    While the back-gate shifts the entire potential of the whole structure,
    the top-gate bridge overlaps with a 200~nm part of the CNT
    by an oxidized Cr layer, see \suppmat{}.
    \panellabel{b}
    Scanning electron-microscopy image of a partly suspended CNT sample.
    The CNT is just visible and indicated by red arrows.
    \panellabel{c}
    Source-drain current through the quantum dot at zero magnetic field as function
    of the bias ($V_\mathrm{sd}$) and top-gate voltage ($V_\mathrm{tg}$),
    adjusting the back-gate voltage $V_\mathrm{bg}$ simultaneously to keep the average chemical potential in the leads constant:
    $V_\mathrm{bg} = 4.35~\mathrm{V}-0.7 \times V_\mathrm{tg}$.
    \panellabel{d}
    $dI/dV_\mathrm{sd}$ at zero magnetic field centered around the $1 \leftrightarrow 2$ single-electron tunneling regime
    for back-gate voltage $V_\mathrm{bg} = 4.25~\mathrm{V}$ in the hole regime.
    We count the number of electrons relative to the last filled conduction band shell of the CNT as usual.
    The diagonal dashed lines marked A-F correspond to transitions between the $N=1$ and $2$ electron quantum-dot states.
    \panellabel{e}
    Energy diagram of the one- and two-electron quantum dot states involving the first orbital shell and the corresponding orbital fillings discussed in the text.
    These states are responsible for the transitions
    A ($D_-  \leftrightarrow S_-$),
    B ($D_-  \leftrightarrow T_0$) and
    C ($D_-  \leftrightarrow S_{0}$)
    which are the most relevant ones for the present discussion, see also \suppmat{}.
    For \Panel{d} we extract $\Delta = 0.8~\mathrm{meV}$ and $J = 1.2~\mathrm{meV}$.
    \panellabel{f}
    Measured magnetic field dependence of the electronic excitation
    lines along the horizontal line in \Panel{d} at $V_\mathrm{sd} = 6~\mathrm{mV}$.
    This way of plotting (see \suppmat{})  focuses the attention on the important triplet states by making the transition B into the lowest triplet appear as a vertical line.
  }  
  \label{fig:1}
\end{figure*}

\begin{figure*}[\figopt]
  \includegraphics[width=0.95\linewidth]{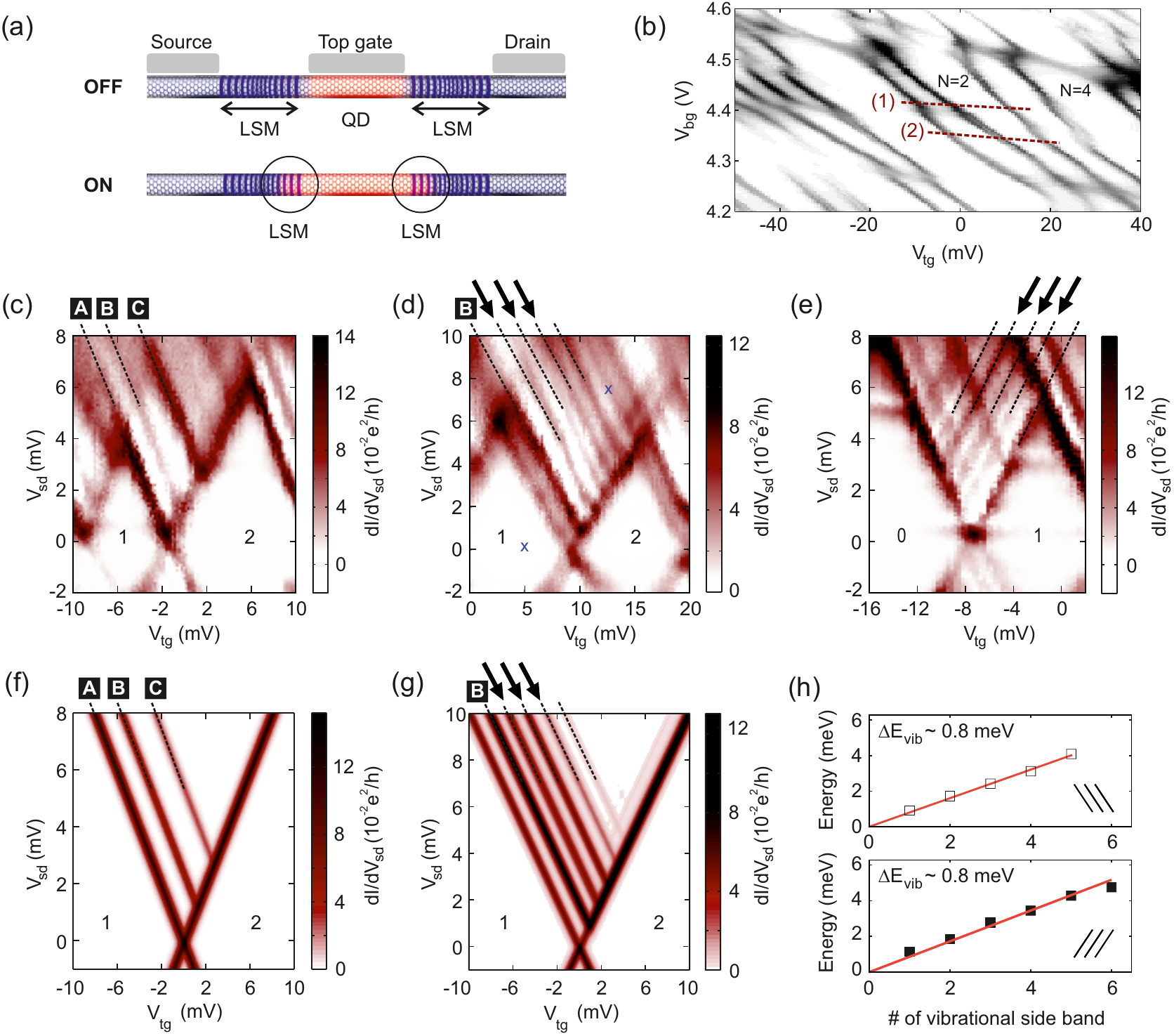}
  \caption{
    Switching the coupling to the vibration {``on''} and{``off''}.
    \panellabel{a}
    Schematic illustration of the quantum-dot tuning
    into a region with a longitudinal stretching mode (LSM).
    \panellabel{b}
    Top- and back-gate voltage stability diagram recorded for
    $V_{\mathrm{sd}} = 1~\mathrm{mV}$.
    The lines marked
    (1) [$V_\mathrm{bg} = 4.41 ~\mathrm{V} - 0.5 \times V_\mathrm{tg}$] and
    (2) [$V_\mathrm{bg} = 4.35 ~\mathrm{V} - 1.0 \times V_\mathrm{tg}$]
    indicate different regimes of electron-vibration coupling.
    \panellabel{c}
    $dI/dV_\mathrm{sd}$ measured along line (1) in \Panel{b}
    showing no effects of vibrations.
    \panellabel{d}
    Measurement of $dI/dV_\mathrm{sd}$ along line (2) in \Panel{b},
    where significant electron-vibration coupling is observed:
    the arrows indicate the vibrational sidebands introduced.
    Electronic lines A and B from \Panel{c} can still be identified,
    but C is commensurate with a vibrational sideband of B.
    [Note that the same happens in the calculations in \Panel{g}.]
    The blue markers indicate the end-points of the line (not shown for clarity) along which the measurements in \Fig{fig:3}{} are taken.
    \panellabel{e}
    Similar measurement as in \Panel{d} but for a different relation of gate voltages
      ($V_{\mathrm{tg}}= 4.45 ~\mathrm{V} - 0.95 \times V_{\mathrm{bg}}$)
    showing vibrational excitations (arrows) with different gate-voltage slope, both in magnitude and sign.
    \panellabel{f},\panellabel{g}
      Calculated $dI/dV_\mathrm{sd}$ corresponding to \Panel{c} and \panel{d},
      respectively, see text.
    The overall conductance magnitude is adjusted through the coupling $\Gamma$,
    taking $T = 0.7~\mathrm{K}$.
    \panellabel{h}
    Linear fit of the vibrational excitations:
    the upper panel fits data from \Fig{fig:2}{d} and the lower panel from \Fig{fig:2}{e},
    both confirming a harmonic spectrum with
    $\Delta E_{\mathrm{vib}}= \hbar \omega = 0.8 \pm 0.1~\mathrm{meV}$
    corresponding to $193 \pm 24~\mathrm{GHz}$.
  }
  \label{fig:2}
\end{figure*}

%
%
In \Fig{fig:1}{a} we show a schematic of a typical suspended CNT quantum-dot device whose scanning electron microscope image is shown in \Fig{fig:1}{b}. The CNT is electrically and mechanically connected to both source (s) and drain (d) contacts where the central electrode acts as a suspended, doubly clamped top gate (tg). The quantum dot is formed  in the small band gap CNT by the electrostatic potentials of the top and back gate (bg), see Figures~\fig{fig:1}{a} and \fig{fig:2}{a}, allowing  for electrostatic control of the size of the quantum dot in the range of 250 - 350~nm, see \suppmat{}.
By changing the gate voltages we can modify the position and size of the dot with respect to the suspended vibrating region of the CNT, which is a crucial part of our experiment.
The high quality of our CNT sample is revealed by the observation of well-resolved, multiple four-fold shell-structure of the electronic states in the stability diagram in \Fig{fig:1}{c} measured at zero magnetic field and at a base temperature of 1.6~K.
This shell-structure stems from the combined spin and valley degeneracies in clean CNTs~\cite{Liang02b,Buitelaar02,Sapmaz05b}, and enables a first characterization of the electronic properties by the Coulomb and confinement energies. Importantly, the resulting estimates show that the quantum dot formed in the CNT is comparable to or even larger in size than  the top-gate, see \suppmat{}.

%
%

The key advantage of our device, in contrast to previous ones, is that we can first obtain detailed information about the electronic spectrum by measuring the differential conductance in a gate voltage regime without signatures of vibrational excitations. For example, in the spectrum shown in \Fig{fig:1}{d} the low-energy excitations indicated by dashed black lines can be assigned to transitions between states with electron number $N=1$ and $2$, respectively.
These are indicated in the schematic in \Fig{fig:1}{e}
which shows for $N=1$ two spin doublets denoted $D_{-}$ and $D_{+}$, 
obtained by filling the (anti)bonding orbitals $\ket{\pm} = (\ket{K} \pm \ket{K'})/\sqrt{2}$  of the $K$ and $K'$ valleys with one electron,
which are split in energy by $2\Delta$ due to the valley-mixing $\Delta$.
For $N=2$  we have spin-singlets $S_{-}$ and $S_{+}$ (latter not shown) completely filling \emph{one} of these orbitals, and a singlet $S_{0}$ and a triplet $T_0$ in which two \emph{different} orbitals are filled.
Here the labels of the  many-body states $S$, $D$, $T$ indicate the spin multiplicities (singlet, doublet, triplet), whereas the subscripts indicate the relevant orbital polarizations.
In the transport data of \Fig{fig:1}{d} we identify a ground singlet ($S_{-}$), an excited triplet ($T_0$) and another singlet ($S_{0}$), split by the exchange energy $J$.
The measured magnetic field transport spectroscopy in \Fig{fig:1}{f} confirms this assignment:
the slope of the lines A and C for transitions to $S_{-}$ and $S_0$, respectively, differs by the Zeeman spin splitting from the slope of line B for the transition to the triplet $T_0$.
We note that for these parameters the singlet $S_{+}$  is the highest in energy in \Fig{fig:1}{e}.
It is not shown there nor discussed further below because this state does not influence the measured transport in the considered regime.~\bibnote[Note1]{
In this state two electrons fill the \emph{excited} orbital causing the amplitude for a transition from the $N=1$ ground state with an electron in the \emph{ground} orbital to be strongly suppressed. For this transition to happen, one must shuffle the lower electron up and add another electron, something which is only possible for very strong spin-orbit coupling or  higher-order tunnel processes, neither of which are relevant here.
This point is important since due to polaronic renormalization discussed at the end of the letter $S_{+}$ may actually lie below $S_{0}$, but even then it cannot be observed.}
 Our calculations below do, however, include the state $S_{+}$ and confirm that it has negligible influence.

%
%
By independently tuning the top- and back gate voltages we can change the electrostatic confinement of the quantum dot and thereby effectively operate a single quantum dot system which can be made sensitive to the vibrating part of the CNT, as illustrated in \Fig{fig:2}{a}. The resulting electronic stability diagram in \Fig{fig:2}{b}, showing nearly parallel lines, indicates that we can independently fix the electron number in the dot while modifying its shape, dimensions and position. When measuring the Coulomb diamonds along the lines indicated in \Fig{fig:2}{b} one expects, electronically speaking, no qualitative difference. Indeed, along the initial working line marked as (1) in \Fig{fig:2}{b}, the measurement in \Fig{fig:2}{c} shows no indications of vibrations. However, when tuning to the working line (2), the excitation spectrum, shown in \Fig{fig:2}{d}, changes in a way that cannot be explained by a modification of the size-quantization energy on the quantum dot: for several subsequent charge states a dense spectrum of discrete excitation peaks appears,
equally spaced by $\hbar \omega =0.8 \pm 0.1~\mathrm{meV}$ as \Fig{fig:2}{h} shows. This is the case across the entire electronic shell that we measure, see \suppmat{}. The spacing lies in the range expected for the high frequency of the longitudinal stretching mode (LSM) of the suspended parts of the CNT (length  $\approx$ 65~nm as  in previous studies~\cite{Leturcq09,Sapmaz05}).
Furthermore, the predominance of the excitation lines with negative slope indicates that the quantum dot couples to only one of the two suspended parts~\cite{Cavaliere10a,Donarini12}.
In \Fig{fig:2}{e} we demonstrate  that by tuning to a different voltage regime we are able to make the other vibrating part dominate.
Our system thus displays electrostatically tuneable electron-vibration coupling.

\begin{figure}[\figopt]
  \includegraphics[width=0.75\linewidth]{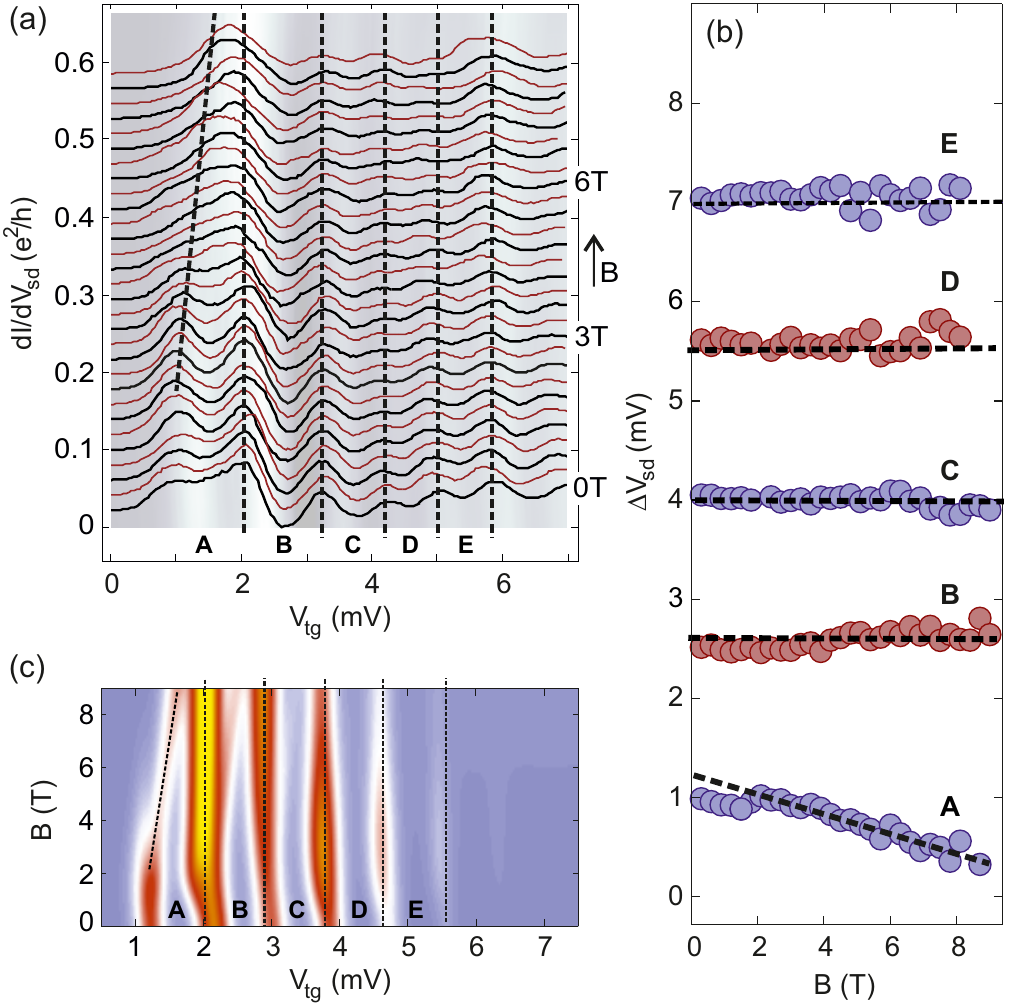}
  \caption{
    {Magnetic field evolution of the vibrational sidebands.}
    \panellabel{a}
    $dI/dV_{\mathrm{sd}}$ as function of top-gate voltage $V_{\mathrm{tg}}$
    along the line (not shown) connecting the blue markers \Fig{fig:2}{d}.
    Each $dI/dV_{\mathrm{sd}}$ line has been measured 
    at a different magnetic field $B$, which has been tuned from $B=0$ to 9~T in steps of 300~mT.
    The $dI/dV_{\mathrm{sd}}$ lines are offset vertically by 0.015~$e^2/h$ for clarity.
    Dashed lines indicate the singlet (sloped) and triplet (vertical)  transitions.
    \panellabel{b}
    Bias-voltage \emph{spacings} ($\Delta V_{\mathrm{sd}}$) of peaks A-E in \Panel{a} as function of the magnetic field
    including additional intermediate line traces that are not shown there.
    The dashed line for the peak-spacing A corresponds to $g$-factor 2.
    The data are offset vertically by 1.5~mV for clarity.
    \panellabel{c}
    Calculated $dI/dV_{\mathrm{sd}}$ evolution corresponding to \Panel{a}.
  }
  \label{fig:3}
\end{figure}

%
%
To illustrate how the switchable coupling to a quantized vibration can be exploited,
we now focus on measurements for the $N=1 \leftrightarrow 2$ electron regime in \Figs{fig:2}{c}{d}. The corresponding calculations shown in \Figs{fig:2}{f}{g} are based on a quantum-dot model including the electronic states identified before in \Figs{fig:1}{d}{e} and coupling to a single vibrational mode. This model will be discussed in detail below, once we have presented all experimental data.
%
%
Apart from this, the non-equilibrium transport is obtained from standard master equations (see \suppmat{}) which incorporate single-electron tunneling into both orbitals of the shell (with asymmetry parameter $\kappa$) from both electrodes (with junction asymmetry parameter $\gamma$).
The electronic and vibrational states are assumed to relax with a phenomenological rate which exceeds the tunneling relaxation rates,
taken for simplicity to be proportional to the energy change $E$ in the transition:
$\Gamma_\mathrm{rel}(E) = \Gamma \times (E/0.2~\mathrm{meV})$.
The overall tunneling rate $\Gamma$ merely sets the scale of the current and is irrelevant to the relative magnitude of the different excitations which is of interest here.

%
%
To experimentally identify the electronic states to which the vibrational excitations belong,
we have investigated  how the differential conductance measured along the line (not shown) connecting the blue markers in \Fig{fig:2}{d}
evolves with a magnetic field $B$ applied perpendicular to the CNT.
The dominating features in \Fig{fig:3}{a} are the vibrational sidebands of the lowest of the triplet excitations $T_0$ which in this presentation of the data appear as vertical lines. Strikingly, the ground state singlet $S_{-}$ evolving with a slope has no vibrational sidebands as demonstrated by fits of the difference of the peak position in \Fig{fig:3}{b}.
This can not be explained by an Anderson-Holstein model where all electronic states with the same charge couple equally to the vibration, see \suppmat{} for explicit attempts.

\begin{figure}[\figopt]
  \includegraphics[width=0.5\linewidth]{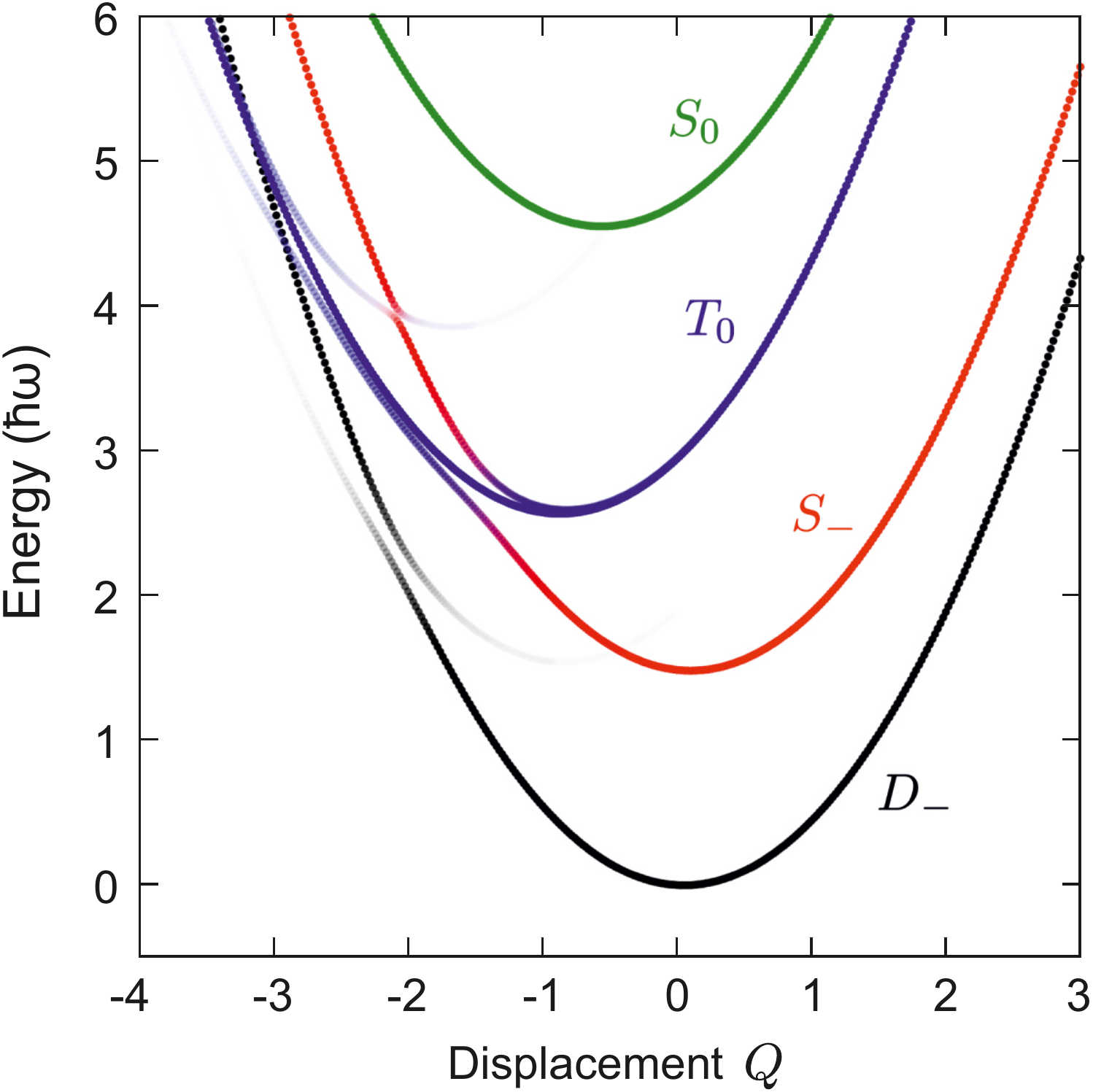}
  \caption{
    Vibrational potential energy surfaces underlying the quantum states of the vibrating CNT quantum dot
    included in the transport calculations using the same parameters as in \Figs{fig:2}{f}{g} and \fig{fig:3}{c}.
    The excited triplet $T_0$ and ground singlet $S_{-}$ have significantly different couplings to the vibration,
    i.e., shifts of their potential minima relative to that of the one-electron ground state $D_{-}$
    strongly differ.
    Due to the weak SO coupling several avoided crossings can be seen.
    The most important anticrossing is that of the $T_0$ (blue) and $S_{-}$ (red) potential energy surfaces.    
    This can be understood directly from the SO operator as written in the text:
    it "flips" both the orbital ($\tau$) and spin index ($\sigma$) of an electron.
    \New{
     In \Fig{fig:1}{e} this implies that for $N=2$ the red spin up in the higher orbital
     is flipped into the blue spin down in the lower orbital
     (this represents a flip from $T_0$ to $S_{-}$).
    The resulting admixture of $T_0$-components (blue) to $S_{-}$ (red)
    causes the latter to remain visible in the transport in \Fig{fig:3}{c}
    with increasing the magnetic field when the tunneling becomes spin-selective due to the CNT leads.
    The remaining SO anticrossings are discussed in the \suppmat{},
    which for our parameters have negligible impact on transport.
  }
  }  
  \label{fig:4}
\end{figure}

%
%
Instead, in our modeling we must account for \emph{state-dependent} Franck-Condon shifts resulting in the vibrational potentials plotted in \Fig{fig:4}{}.
To arrive at this, we start from a model accounting for the observed set of accessible~\cite{Note1} many-body transport states, which is restricted by Coulomb blockade and bias voltage of a few mV to those shown in \Fig{fig:1}{e}  with electron numbers $N=1$ and $N=2$ and a single electronic $K$-$K'$ shell:
\begin{equation}
  H_\mathrm{el}
  =
  \varepsilon N
  + \Delta \sum_{\tau=\pm}\tau \sum_\sigma  d_{\tau \sigma}^\dagger  d_{\tau \sigma}
  -J\,\mathbf{S}_{+}\cdot \mathbf{S}_{-}
  .
  \label{eq:ham}
\end{equation}
Here $\varepsilon$ is mean level position controlled by $V_{\mathrm{tg}}$,
$\Delta$ is the subband or valley-mixing term
and $J$ is the exchange coupling between the spins in the two orbitals $\tau=\pm$
with spin-operators $\mathbf{S}_{\tau} = \tfrac{1}{2} \sum_{\sigma,\sigma'} d^\dag_{\tau\sigma} \boldsymbol{\sigma}_{\sigma,\sigma'}d_{\tau\sigma'}$
[$\sigma,\sigma'$ are spin indices, $\boldsymbol{\sigma}$ is the usual vector of Pauli-matrices,
and $d_{\tau \sigma}^\dag$ creates a spin-$\sigma$ electron in orbital $\tau$].
%
%
To obtain a result as plotted in \Figs{fig:2}{f}{g} we first introduced a Holstein coupling by allowing the level position $\varepsilon$ to depend on $Q$, the dimensionless vibration coordinate normalized to the zero-point motion:  we thus formally replace
$\varepsilon \rightarrow \varepsilon +  \sqrt{2} \hbar \omega \lambda_{\varepsilon} Q$.
This results in the commonly assumed {uniform} vibration coupling with strength $\lambda_{\varepsilon}$ to all electronic states with the same charge $N$, which is not observed here.
The required {state dependent} electron-vibration coupling is obtained by additionally accounting for a dependence of the other parameters on the vibration coordinate, i.e.,
we formally replace
$\Delta \rightarrow \Delta+ \sqrt{2} \hbar \omega \lambda_{\Delta} Q$,
where $\lambda_{\Delta}$ is a dimensionless one-electron {valley-vibration} coupling,
and 
 $J \rightarrow J+ \sqrt{2} \hbar \omega \lambda_J Q$,
where $\lambda_J$ is a dimensionless coupling of the vibration to the two-electron exchange.
Here many-body physics  comes in:
when going from the singlet $S_{-}$ ground state to the triplet $T_0$, the Pauli principle forces the two electrons into \emph{different} orbitals which can couple differently to the vibrational mode (difference quantified by $\lambda_\Delta$).
However, the coupling $\lambda_J$ is important as well:
 when allowing only for $\lambda_\Delta$, the effective electronic excitation spectrum  for fixed charge $N$ (relative to which the vibration excitations are ``counted'') becomes dependent on the vibrational couplings (polaronic renormalization).
That experimentally no significant shift of the electronic excitations is found  when turning {``on''} the couplings to the vibration
requires the couplings $\lambda_J$ and $\lambda_\Delta$ to be comparable in magnitude but opposite in sign. This results in an enhanced coupling of the triplet $T_{0}$ over $S_{-}$ while the polaronic shifts that they induce cancel out, keeping the effective electronic excitations fixed.
This thus leaves one free parameter, their magnitude, which controls the degree of state-specific coupling, which we adjust to the experiment.
Together this suffices to obtain results such as \Figs{fig:2}{f}{g} that reproduce the main zero-field observations of \Figs{fig:2}{c}{d}.
When the vibration couplings are ``off' in \Fig{fig:2}{f} we estimate from \Fig{fig:2}{c} the parameter values
$\Delta = 0.8~\mathrm{meV}$,
$J = 1.5~\mathrm{meV}$,
(similar to those in \Fig{fig:1}{d})
and use $\gamma =0.0$,
$\kappa = -0.3$.
When the vibration couplings are ``on'' in \Fig{fig:2}{g}
we use the same values for $J$ and $\Delta$ but nonzero vibration couplings
$\lambda_\varepsilon=0.28$,
$\lambda_\Delta=0.32$,
$\lambda_J=-0.22$
and frequency $\hbar \omega = 0.85~\mathrm{meV}$
and we adjusted the asymmetries
$\gamma =-0.5$,
$\kappa = 0.3$.
Despite the fact that there are several parameters,
the experiment imposes strong restrictions, in particular,
regarding the choice of vibrational couplings,
excluding a simple Holstein mechanism  ($\lambda_\Delta= \lambda_J=0$),
see \suppmat{}.
We arrive at the three electron-vibration couplings by imposing three experimental constraints after
expressing the effective couplings of the electronic states in terms of $\lambda_\varepsilon$, $\lambda_\Delta$, and $\lambda_J$:
(i) the observed $T_0$-$S_{-}$ splitting and (ii) $S_0$ - $T_0$  splitting (commensurate with $2 \hbar \omega$) should match energy expressions that depend on the vibrational couplings (polaron shift)
and (iii) the vibrational-coupling of $T_0$ is adjusted to numerically reproduce the observed number of triplet vibrational sidebands.
We note that in \Figs{fig:2}{d}{e}, the higher vibrational sidebands become more intense at high bias. As expected, this is not captured by our model since this may involve excitations beyond the lowest two electronic orbitals and  energy-dependence of the tunnel barrier, neither of which we include here. We have focused instead on the nontrivial interplay of vibrational and spin-excitations for $N=1$ and $N=2$ in the lowest sidebands.

The resulting physical mechanism is illustrated in \Fig{fig:4}{}:
when starting out from state $D_{-}$ and adding a second electron to the lowest orbital the lowest singlet state $S_{-}$ experiences only a small horizontal shift of the vibrational potential minimum (both electrons in orbital $\ket{-}$ have their coupling weakened by $\lambda_\Delta$ and there is no spin and therefore no exchange modification of the coupling by $\lambda_J$).
However, when adding the electron to the excited orbital, the coupling is not only enhanced by $\lambda_\Delta$, but also by a negative $\lambda_J$ when a spinfull triplet $T_0$ is formed. This results in a large Franck-Condon shift of the potential minimum of $T_{0}$ in \Fig{fig:4}{}.
The above horizontal shifts of the potential minima translate into suppressed vibrational sidebands for the singlet $S_{-}$ and a pronounced series of sidebands for the triplet $T_0$, respectively (Franck-Condon effect).
The presence of the further electronic states and their quantized vibrational states in \Fig{fig:4}{}, all of which are included in our transport calculations,
do not alter the above simple picture:
Whereas the excited singlet $S_{+}$ does not couple to transport~\cite{Note1}, the role of $S_{0}$ cannot be ascertained at zero magnetic field because it is commensurate (within the line broadening) with one of the vibrational sidebands of $T_0$.
%
%

The field evolution in \Fig{fig:3}{c}, calculated by adding a Zeeman term to \Eq{eq:ham}, reproduces the  main observation of \Fig{fig:3}{a}, namely, that the triplet maintains its vibrational sidebands (vertical) but the \emph{ground} singlet $S_{-}$ (sloped) does not. However, to obtain this agreement with the measurements we are forced to further extend the above model.
%
%
First, both the \emph{excited} singlet ($S_{0}$) as well as the Zeeman split-off states of the triplet ($T_0$) do not appear in the measurements.
This we attribute to the fact that the source and drain leads of the quantum dot are not formed by metallic contacts but by small pieces of suspended CNT.
Zeeman splitting of discrete states in these CNT contacts may lead to spin-filtering which turns on with the magnetic field,
developing full strength at a few Tesla where $g \mu_{\mathrm{B}} B \approx k_{\mathrm{B}} T$.
We phenomenologically account for this by a spin-dependence in tunneling to / from the electrodes which depends on $B$: $\zeta(B) = \tanh (g\mu_{\text{B}} B/2k_{\text{B}} T)$.
%
%
Second, when only including this spin-filtering in the model, it suppresses the singlet groundstate $S_{-}$ (without vibrational bands) which we do experimentally observe as excitation A in \Figs{fig:3}{a}{b}.
However, when even a small spin-orbit (SO) coupling is included,
the singlet $S_{-}$ reappears (borrowing intensity from the triplet $T_{0}$, cf. also \Fig{fig:4}{}), but, importantly, without reinstating the unobserved $S_{0}$ and the Zeeman split-off states of $T_0$ and their vibrational sidebands.
This produces the observed intensity pattern, which is impossible to achieve with simple commonly used models, see \suppmat{}.
Here, the spin-orbit coupling is included by adding to \Eq{eq:ham}
a term $H_\mathrm{SO}= \Delta_\mathrm{SO}\sum_{\sigma,\tau} d^\dag_{\tau\sigma}  d_{-\tau-\sigma}$
with $\Delta_\mathrm{SO}=0.1~\mathrm{meV}$ which allows both the spin $\sigma$ and orbital index $\tau$ to be flipped in the schematic \Fig{fig:1}{d}, thereby coupling in particular $T_0$ to $S_{-}$, lending it intensity.
\Figs{fig:2}{f}{g} and \fig{fig:3}{c} are  based on the inclusion of all these effects.
However, we emphasize, that in the latter figure spin-filtering and spin-orbit coupling are needed exclusively to explain the missing Zeeman lines, but do not lead to a qualitative change of the state-dependent coupling at $B=0$ in \Fig{fig:2}{g}, which is our main finding.
The \suppmat{} explores the influence of the various parameters, confirming the necessity of including them.
The key advantage of our tuneable setup is that we are able to first identify excitation A and B as relating to electronic singlet $S_{-}$ and triplet $T_0$, respectively, and subsequently allowing us to study the vibrational sidebands C-E.

In conclusion, we have demonstrated switchable coupling of a quantized vibration of a carbon nanotube to its quantized electronic states.
Using this advance we explored the two-electron regime -- including the magnetic field dependence -- and found indications of state-dependent vibrational transport sidebands not described by standard models.
We showed that the interplay of intrinsic effects on the carbon nanotube (Coulomb blockade, valley-index, spin-exchange) and experimental details (junction, orbital, and spin asymmetries) can explain the observations.
This, however, includes vibrational couplings that involve internal spin- and valley-degrees of freedom, bringing spin- and valley-tronics physics within range of NEMS.

\subsection*{Author information}
Corresponding author:
stampfer@physik.rwth-aachen.de.
\\
Notes:
The authors declare no competing financial interests.

\begin{acknowledgement}
We acknowledge F. Cavaliere for stimulating discussions,
and S. Trellenkamp, J. Dauber for support with sample fabrication.
We acknowledge support from the Helmholtz Nanoelectronic Facility (HNF)
and financial support by the JARA Seed Fund and the DFG under Contract No. SPP-1243 and FOR912.
\end{acknowledgement}

\begin{suppinfo}
Fabrication and experimental characterization of the quantum dot, the electrostatic control of the coupling to vibrational modes and
 theoretical analysis of  the electronic and vibrational quantum states of the model and the transport calculations using master equations.
\end{suppinfo}

\providecommand*\mcitethebibliography{\thebibliography}
\csname @ifundefined\endcsname{endmcitethebibliography}
  {\let\endmcitethebibliography\endthebibliography}{}



\section*{Supplementary material}

# Supplementary Information for

# 'Switchable coupling of vibrations to two-electron carbon-nanotube quantum dot states'

P. Weber,[1, 2] H. L. Calvo,[3, 2] J. Bohle,[3, 2] K. Goß,[4, 2] C. Meyer,[4, 2] M. R. Wegewijs,[3, 4, 2] and C. Stampfer[1, 4, 2]

[1] *2nd Institute of Physics, RWTH Aachen University, 52056 Aachen, Germany*
[2] *JARA – Fundamentals of Future Information Technology**
[3] *Institute for Theory of Statistical Physics,*
*RWTH Aachen, 52056 Aachen, Germany*
[4] *Peter Grünberg Institut, Forschungszentrum Jülich, 52425 Jülich, Germany*

## CONTENTS

In this Supplementary Information we provide a detailed description of the experimental methods and additional measurements (Sec. I) and a precise formulation of the model and calculations reported in the main text (Sec. II). In both sections, we provide an extensive discussion of claims and results of the main article. Within the Supporting Information references are numbered as, e.g., equation (S-1) and Figure S-1, whereas regular numbers, e.g., equation (1) and Figure 1, refer to the main article.

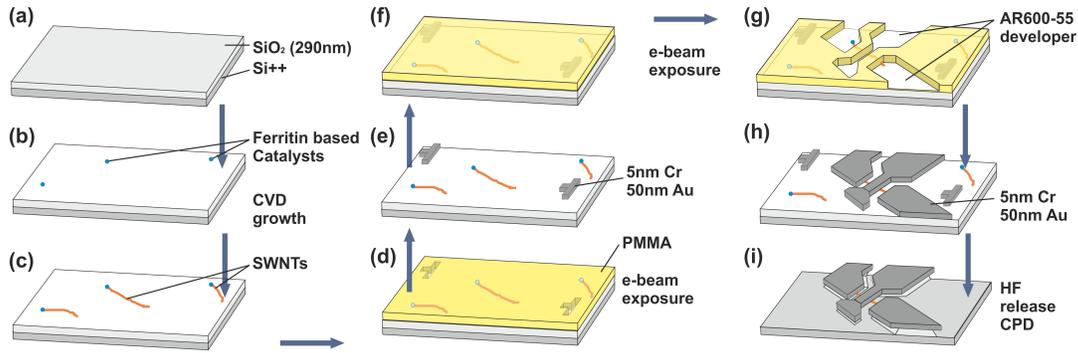

FIG. S-1. Fabrication work flow described in the text.

# I. EXPERIMENTAL METHODS

## A. Fabrication

Devices were fabricated in a similar fashion as in Ref. [1] as outlined in Fig. S-1. The starting point is a highly doped silicon wafer covered by 290 nm silicon oxide (step a). Next, following the recipe described in Ref. [2] ferritin catalyst nanoparticles are dispersed on the substrate (step b) from which carbon nanotubes are grown by means of chemical vapor deposition (step c). For the subsequent selection and localization of carbon nanotubes marker structures are evaporated in an electron beam (e-beam) lithography step (step d,e). In a second e-beam lithography step metallic electrodes and gate structures are deposited in a single evaporation (5nm Cr, 50nm Au) on selected carbon nanotubes (step f-h). Finally, diluted hydrofluoric acid (1% for 6min) is used to etch the silicon oxide followed by critical point drying (step i). It is crucial that the central electrode is completely underetched so that the chromium layer oxidizes when exposed to environmental air to form the top-gate oxide.

## B. Configuration of the quantum dot

In Fig. S-2 we show the back-gate voltage ($V_{bg}$) characteristics of the investigated carbon nanotube (CNT) at two different temperatures. The CNT is slightly p-doped and the charge neutrality point is found to be around 7 V. At gate voltages close to the charge neutrality point the current $I_{sd}$ is strongly suppressed, even for elevated temperatures $T = 50$ K. This is the typical characteristics of a CNT with a small semiconducting gap separating p- and n-conducting regions. For low temperatures the current within the semiconducting band gap, apart from a few small resonances, is pinched off. The band gap extends over a back-gate voltage range of $\Delta V_{bg} = 1.5$ V. Multiplied with the absolute lever arm of the back gate $\alpha_{bg} = 0.035$ the energy splitting between valence and conduction band is $\Delta E_{gap} \approx 50$ meV. At $T = 1.6$ K we observe a number of reproducible resonances. These resonances are attributed to Coulomb blockade effects. The irregularity of the resonances indicates the existence of multiple quantum dots and a non-monotonic CNT band structure along the nanotube axis.

In Fig. S-3a we illustrate the electrostatic formation of the main quantum dot. The Fermi level along the entire CNT can be tuned with respect to valence and conduction band by an applied back-gate voltage $V_{bg}$. Owing to the geometry of the device (see Figures 1a - 1b of the main text) different electrostatic potentials act on different parts of the nanotube depending on the proximity to the local top gate: The electrons just below the top gate experience electrostatic screening of the back-gate voltage because the CNT is separated only by a few nanometers of oxide from the top-gate electrode while a combination of top gate and back-gate voltage is acting on the suspended parts of the CNT. Therefore, we expect a bending of the CNT band structure along its axis.

A device schematics together with an indication of the location and size of the quantum dot with respect to the metallic leads and the top gate is given in Fig. S-3b. As deduced below, the quantum dot is formed close to the top gate and its lateral size is on the order of the top-gate

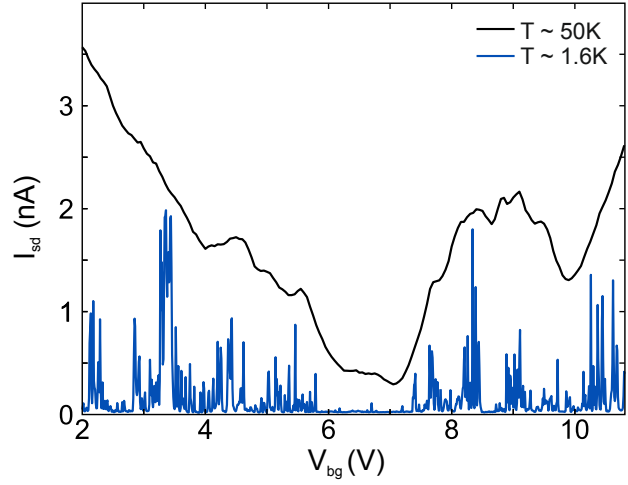

FIG. S-2. Source-drain current $I_{\mathrm{sd}}$ through the CNT as a function of back gate voltage $V_{\mathrm{bg}}$ at small bias voltage $V_{\mathrm{sd}} = 0.3$ mV at two different temperatures.

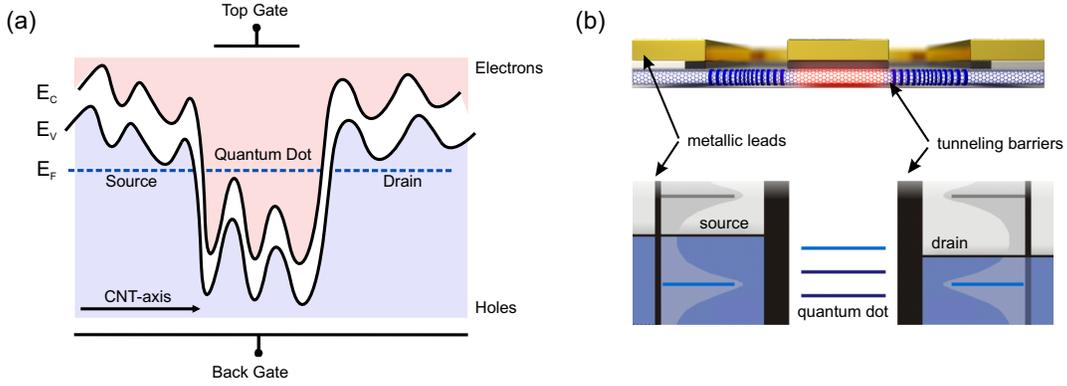

FIG. S-3. Formation scenario of the quantum dot. (a) Electrostatic bending of the CNT bands in close vicinity to the top gate induced by top and back gate voltages. (b) The upper illustration shows a schematic of the device. The semi-suspended CNT is connected to Cr/Au leads and a central top gate. In the suspended region longitudinal stretching modes (LSM) of vibration are indicated by the blue spirals surrounding the CNT, while the red region below the top gate symbolizes the quantum dot. The lower schematics describes the tunneling through the quantum dot. Electrostatically induced tunnel barriers separate the CNT quantum dot electrically from the CNT leads.

width of 200 nm. These parameters suggest, that the leads of the QD are not the metallic contacts but the CNT itself and therefore the size of the QD can be tuned exclusively by the applied gate voltages.

In order to determine the electronic size of the quantum dot we analyse the addition energies needed to add the first and the second electron on the investigated electronic shell of the QD, respectively. The addition energy is defined as the change in electrochemical potential $\Delta\mu_N$ when adding the $(N+1)$ charge to a quantum dot containing already $N$ charges[3,4]. It can be related to the charging energy $E_{\mathrm{C}}$, the quantum energy-level separation $\beta$, the valley degeneracy splitting $2\Delta$ and the exchange interaction $J$ via

$$\Delta\mu_0 = E_C + \beta - 2\Delta \tag{S-1}$$

$$\Delta\mu_1 = E_C. \tag{S-2}$$

Additionally, we can infer $J$ and $\Delta$ from the energies of the first two excited states relative to the two-electron ground state, $E_{T_0} - E_{S_-} = 2\Delta - J/4$ and $E_{S_0} - E_{S_-} = 2\Delta + 3J/4$, respectively [cf. equation (S-14), (S-45), and (S-46) below]. From Fig. 2c of the main article [reproduced in Fig. S-6b below] we extract $\Delta\mu_0 = 9.0 \pm 0.3$ meV, $\Delta\mu_1 = 3.8 \pm 0.4$ meV, $\mu_1 = 1.2 \pm 0.3$ meV and

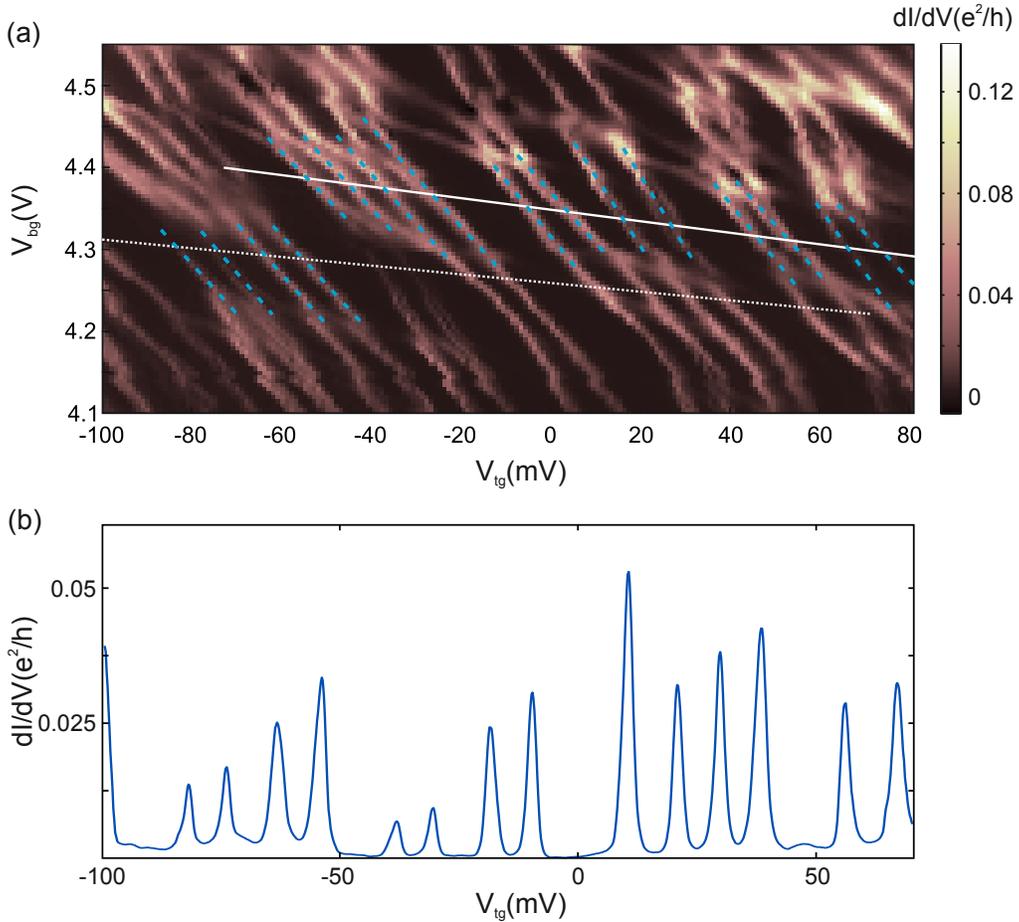

FIG. S-4. (a) Charge stability map in the $V_{tg}$-$V_{bg}$-plane. The differential conductance is plotted for zero d.c. bias voltage $V_{sd}$ applied between source and drain. Resonances with a slope marked by the blue dashed lines are attributed to resonances off the central quantum dot. The Coulomb diamonds in Fig. 1c of the main article are measured along the white continuous line and the Coulomb peaks in (b) are measured along the white dashed line ($V_{bg} = 4.26$ V $- 0.6 \times V_{tg}$).

$\mu_2 = 2.7 \pm 0.3$ meV. This yields

$$\beta = 6.8 \text{ meV}, \quad J = 1.5 \text{ meV} \quad \text{and} \quad \Delta = 0.8 \text{ meV.} \tag{S-3}$$

Due to large level spacing $\beta$ relative to $J$ and $\Delta$, the analysis of the transport spectrum involving the first few vibrational excitations – at the focus of the main article – only requires considering electron fillings of the first orbital shell (i.e., of the orbitals labeled $\tau = \pm$ in the main article). A more detailed fitting of the energy positions consistent with the above and including the vibrations is given in Sec. II C 1. The expression for the quantization-induced level spacing $\beta = h v_F/2L$, with $v_F = 8.1 \times 10^5$ m/s allows to determine the quantum dot length $L = 245 \pm 25$ nm, corresponding approximately to the length below the top gate.

For the determination of the position of the quantum dot we have evaluated the respective lever arms of both top gate and back-gate electrodes as well as the relative lever arms of the source and drain leads. The analysis of the slopes of the edges of the Coulomb diamonds in Fig. 1d of the main article gives the following lever arms: $\alpha_{tg} = 0.55 \pm 0.04$ and $\alpha_s - \alpha_d = 0.01$.

Knowing the lever arm to the top gate, the lever arm to the back gate can be obtained from the charge-stability map when varying both top gate and back-gate voltages, as shown in Fig. S-4. This figure consists of lines with different slopes, which is a signature of a multiple quantum dot structure. The interesting features are the ones with the largest slope, marked by the blue dashed lines. The slope of these lines allows to extract the relative lever arm between top gate and back gate $\alpha_{rel} = \alpha_{tg}/\alpha_{bg} = 15.5 \pm 1.9$, resulting in $\alpha_{bg} = 0.035 \pm 0.005$.

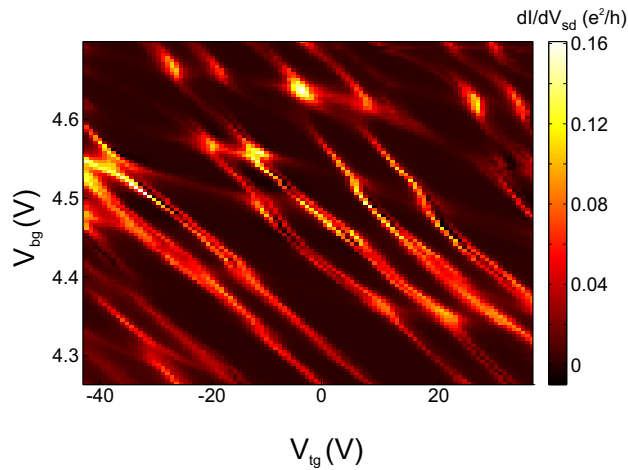

FIG. S-5. Charge stability diagram in both gate voltages for finite d.c. bias voltage of $V_{sd} = 1$ mV. The differential conductance plot shows splitting of the lines belonging to the central quantum dot.

The significantly larger lever arm to the top gate and the almost negligible difference between the lever arms of source and drain confirm our assumption that the quantum dot is located in the middle of the nanotube in close vicinity of the top gate and is as a result strongly screened from the back gate. Additional lines with smaller slope in the charge-stability diagram indicate the existence of additional quantum dots further away from the top gate.

In Fig. S-4a we show a charge stability map as a function of both the top gate and the back-gate voltage. Strikingly, the Coulomb-peak excitation lines with the steepest slope, which correspond to the central QD, appear in groups of four, reflecting the twofold orbital-degenerate bandstructure of high-quality CNTs. Further confirmation of the fourfold periodicity is provided in Fig. S-4b where we plot a Coulomb-peak measurement as function of $V_{tg}$. These characteristics justify the treatment of the CNT QD as an effective few-electron system.

The CNT-QD is thus connected to the metallic electrodes by short CNT leads. To prove that the CNT-leads do not electrically influence our measurements on the central QD significantly, we have measured the same diagram as in Fig. S-4a, but with a finite bias voltage $V_{SD} = 1$ mV. Fig. S-5 shows that only the lines corresponding to the central QD split up into two lines, while the lines corresponding to the lateral parts are not affected. This broadening into ground and excited states proves that the voltage drop occurs only at the tunnel barriers (marked in Fig. S-3b) between the CNT-QD and the CNT leads. The tunneling rates are estimated to be on the order of $\Gamma = 2\pi \times 10$ GHz, which corresponds to roughly 300-400 mK, well below the experimental temperature $T = 1.6$ K. As a first approximation it thus makes sense to apply a standard master-equation description of single-electron tunneling transport, see Sec. II.

### C. Electrostatic control of the coupling to vibrational modes

In this section we provide further experimental data on the tuning of the coupling of the QD to vibrational modes. In particular, in Fig. S-6 we provide an additional QD excitation spectrum to further illustrate our ability to continuously tune the electron-vibration coupling. Figures S-6a, S-6b and S-6d correspond to Figures 2b, 2c and 2d of the main article and show a pure electronic excitation spectrum and a vibrational excitation spectrum, respectively. In addition, we show in Fig. S-6c data for an intermediate state of the QD.

In order to understand the switchable coupling of the QD to vibrational modes we compare the quantum dot size in the two extreme regimes. Unfortunately, when the vibrations are switched on we can not assign the energy of the second excited state $S_0$ in the two electron regime (because it is degenerate with a vibrational sideband within the experimental line width). Nevertheless, we can give bounds for the dot size. With $\Delta\mu_0 = 9.3$ meV, $\Delta\mu_1 = 5.8$ meV and $\mu_1 = 0.9$ meV (see Fig. S-6d and also Fig. S-9b) we get $\beta = 4.4$ meV $+ J/4$, which gives a larger quantum dot size in the vibrational regime for $0 < J < 9.6$ meV. For realistic exchange energies this inequality

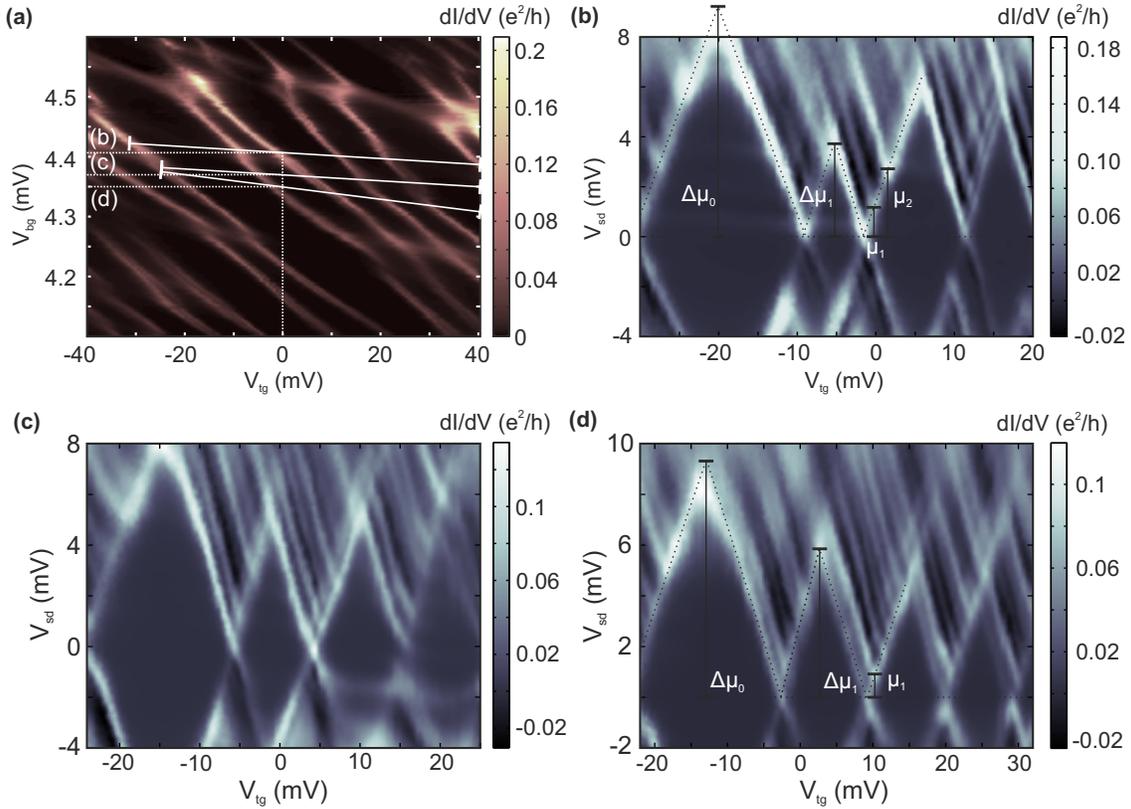

FIG. S-6. (a) Differential conductance in the $V_{\mathrm{tg}}$-$V_{\mathrm{bg}}$-plane for small d.c. bias voltage. (b)-(d) Coulomb diamond charge stability diagrams in the $V_{\mathrm{sd}}$-$V_{\mathrm{tg}}$-plane for the effective charge states $0-4$ in the quantum dot measured along the respective lines in panel (a).

is fullfilled. If we choose, for example, $J = 1.5$ meV as in the regime without vibrations the quantum dot length exceeds 300 nm. The larger quantum dot size, when vibrations are switched on, indicates that the overlap between QD and suspended regions of the CNT is increased. This, in turn, results in coupling to vibrational modes.

## D. Temperature dependence of vibrational sidebands

We also investigated temperature dependence of ground and excited states in the temperature regime where Coulomb blockade peaks still could be resolved along the lines of Ref. [1]. In Fig. S-7a we show Coulomb diamonds measured in the very same region of gate voltage (same electronic state) for different temperatures, 1.6 K, 2.5 K, 3.5 K and 5 K. As the temperature increases, the conductance peaks related to ground and excited states wash out, i.e., they broaden and the conductance maximum decreases. In Fig. S-7b we show the temperature dependence of the maximum conductance $G_{\mathrm{max}}$ for the electronic triplet excited state $T_0$ (blue circles) and its first vibrational replica, i.e. the emission side-band (red triangles). In the four panels in Fig. S-7a these excitations are marked by blue and red arrows, respectively. For the tunneling through the electronic triplet $T_0$ excited state and its vibrational emission peak, we observe a $G_{\mathrm{max}} \sim 1/k_B T$ dependence (blue and red curves) that one expects for the derivative of the Fermi distribution in the quantum Coulomb blockade regime[3].

Unlike in Figs. 2c-2d in Ref. [1] we do not observe any extra vibrational-*absorption* conductance peaks appearing with increasing temperature inside the Coulomb blockaded region, most likely because of weaker electron-vibron coupling strength / smaller tunneling rates. This prohibits further investigation of the temperature dependence.

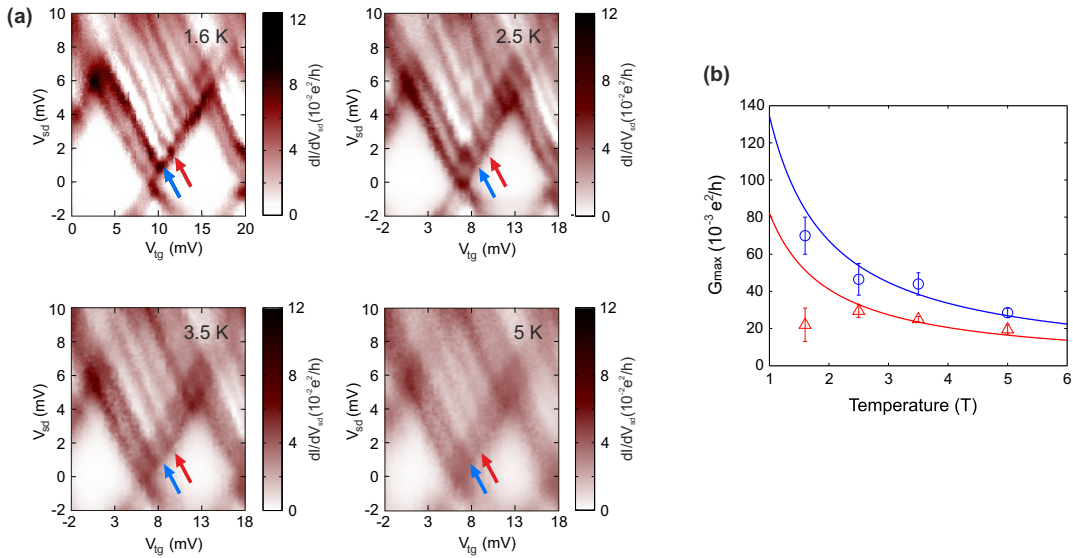

FIG. S-7.  (a) Coulomb diamond charge stability diagrams in the $V_{sd}$-$V_{tg}$-plane around the $N = 1$-$2$ transition for $T = 1.6$ K, as shown in Fig. 2d of the main article, and for three higher temperatures $T = 2.5$ K, 3.5 K and 5 K. (b) Conductance peak value as function of temperature for electronic triplet state (blue circles) and the first vibrational replica (red triangles). The blue and red curves are guides to the eye $\propto 1/k_B T$ .

### E.  Measurements in a magnetic field

The motivation for plotting the data as done in Fig. 3a in the main article is explained in Sec. II C 2.

## II.  THEORETICAL MODELING

In this section we describe in detail the employed model, the method used for calculating the differential conductance of the CNT quantum dot, and the resulting understanding of the transport measurements. The full Hamiltonian of the system under consideration reads as $\mathcal{H} = H_{qd} + H_{tun} + H_{res}$, where $H_{qd}$ describes the quantum dot states, including both their electronic and vibrational degrees of freedom, and $H_{tun}$ is the tunnel coupling Hamiltonian between the dot and the reservoirs described by $H_{res}$.

### A.  Model and eigenstates

#### 1.  Electronic model

*a.  Carbon-nanotube quantum dot*  We first set up an electronic model that accounts for the many-electron states observed in the experiment when the coupling to the CNT vibrations is switched "off". In Sec. II A 2 we then include the vibrations and their coupling to obtain our full CNT quantum-dot Hamiltonian $H_{qd}$. As mentioned at the end of Sec. I B we can restrict our attention to a single orbital shell. We account for a significant valley-mixing $\Delta > 0$:

$$\Delta \sum_{\sigma} (d^{\dagger}_{K\sigma} d_{K'\sigma} + \text{H.c.}),\tag{S-4}$$

where $d^{\dagger}_{K\sigma}$ $(d_{K'\sigma})$ is a creation (annihilation) operator for an electron in valley $K$ $(K')$ with spin $\sigma = \{\uparrow, \downarrow\} = \{1, -1\}$ whose quantization axis is chosen along the direction of the applied magnetic field. Due to the large splitting $\Delta$, it is reasonable to use a basis of bonding $(\tau = -1)$ and

antibonding ($\tau = +1$) combinations (BA) of the $K$ and $K'$ valleys, i.e. $d_{\tau\sigma}^{(\dagger)} = (d_{K\sigma}^{(\dagger)} + \tau d_{K'\sigma}^{(\dagger)})/\sqrt{2}$. In this basis, the electronic model Hamiltonian can be written as

$$H_{\text{el}} = \varepsilon N - g\mu_B B S_z + \Delta \sum_\tau \tau N_\tau - J \mathbf{S}_+ \cdot \mathbf{S}_- + \frac{E_C}{2} N(N-1). \tag{S-5}$$

In the first term, $\varepsilon = -\alpha_{\text{tg}} V_{\text{tg}}$ is the quantum-dot energy level, electrostatically controlled by the top-gate, and $N = \sum_{\tau\sigma} d_{\tau\sigma}^\dagger d_{\tau\sigma}$ is the occupation number operator in the dot. Note that in the experiment back- and top-gate are tuned simultaneously in a linearly dependent way and that $V_{\text{tg}}$ is taken as the independent parameter. Here $N = 1 - 4$ counts the electrons that fill up the orbital shell that we consider and $N = 0$ corresponds to the 'empty dot' state $|0\rangle$ in which all lower shells are filled. The further number operators $N_\tau = \sum_\sigma d_{\tau\sigma}^\dagger d_{\tau\sigma}$ and $N_\sigma = \sum_\tau d_{\tau\sigma}^\dagger d_{\tau\sigma}$ count the number of electrons in orbital $\tau$ and with spin $\sigma$, respectively. The second term in equation (S-5) describes the spin Zeeman splitting due to magnetic field $B$ applied perpendicular to the CNT axis and $S_z = \sum_\sigma (\sigma/2) N_\sigma$ is the operator of the spin component along the field $B$. The third term in the electronic Hamiltonian (S-5) is a spin-exchange term with energy $J$. The spin operator $\mathbf{S}_\tau$ for electrons in orbital $\tau = \pm$ in the BA-basis is $\mathbf{S}_\tau = \sum_{\sigma\sigma'} \boldsymbol{\sigma}_{\sigma\sigma'} d_{\tau\sigma}^\dagger d_{\tau\sigma'}/2$, where $\boldsymbol{\sigma}$ is the vector of Pauli matrices. Finally, the last term is the charging energy $E_C$ accounting for Coulomb repulsion. The inter- and intra-valley electronic repulsion energies are assumed to be the same as is typically observed in CNTs quantum-dot samples[5]. Since we will restrict ourselves to the analysis of the $1 \leftrightarrow 2$ charge regions, both the charging energy $E_C$ and the level energy $\beta$ [cf. equation (S-1)] can be absorbed by a redefinition of the origin of top-gate voltage and therefore these do not need to be included henceforth. Thus setting $E_C = 0$ we obtain the Hamiltonian (1) of the main article, where we note that for brevity the Zeeman term was not written but only mentioned in the main article.

A convenient many-particle basis for $N = 1$ and $N = 2$ charge sectors is constructed by creating spin-multiplets using these orbitals. In the following we will denote by $|x, y\rangle$ such electron fillings, where the first (second) slot $x$ ($y$) corresponds to the ground orbital $-$ (excited orbital $+$). By filling the empty dot state $|0\rangle$ with one electron, we obtain two spin-doublets denoted by $D_\pm$ in the main article, with states

$$|D_-^\sigma\rangle = d_{-\sigma}^\dagger |0\rangle = |\sigma, \bullet\rangle, \tag{S-6}$$

$$|D_+^\sigma\rangle = d_{+\sigma}^\dagger |0\rangle = |\bullet, \sigma\rangle. \tag{S-7}$$

where $\bullet$ denotes an empty $\pm$ orbital, respectively. The next step is to add a further electron to these states. We therefore obtain two 'localized' singlet-fillings of the *same* orbital

$$|S_-\rangle = d_{-\downarrow}^\dagger d_{-\uparrow}^\dagger |0\rangle = |\downarrow\uparrow, \bullet\rangle, \tag{S-8}$$

$$|S_+\rangle = d_{+\downarrow}^\dagger d_{+\uparrow}^\dagger |0\rangle = |\bullet, \downarrow\uparrow\rangle. \tag{S-9}$$

By filling the empty dot with two electrons in *different* orbitals and diagonalizing the electronic Hamiltonian of equation (S-5), we obtain a 'delocalized' singlet

$$|S_0\rangle = \frac{d_{+\downarrow}^\dagger d_{-\uparrow}^\dagger - d_{+\uparrow}^\dagger d_{-\downarrow}^\dagger}{\sqrt{2}} |0\rangle = \frac{|\uparrow, \downarrow\rangle - |\downarrow, \uparrow\rangle}{\sqrt{2}}, \tag{S-10}$$

and a triplet $T_0$ of states $|T_0^m\rangle$ with spin-projections $m = 0, \pm 1$:

$$|T_0^0\rangle = \frac{d_{+\downarrow}^\dagger d_{-\uparrow}^\dagger + d_{+\uparrow}^\dagger d_{-\downarrow}^\dagger}{\sqrt{2}} |0\rangle = \frac{|\uparrow, \downarrow\rangle + |\downarrow, \uparrow\rangle}{\sqrt{2}}, \tag{S-11}$$

$$|T_0^+\rangle = d_{+\uparrow}^\dagger d_{-\uparrow}^\dagger |0\rangle \qquad\qquad = |\uparrow, \uparrow\rangle, \tag{S-12}$$

$$|T_0^-\rangle = d_{+\downarrow}^\dagger d_{-\downarrow}^\dagger |0\rangle \qquad\qquad = |\downarrow, \downarrow\rangle. \tag{S-13}$$

The labels of the many-body states $S$, $D$, $T$ indicate the spin multiplicities (singlet, doublet, triplet), whereas the subscripts indicate the relevant orbital polarizations (signature of difference of number of $\tau = \pm$ electrons, respectively), which is important here.

The above defined states are exact many-particle eigenstates of the model (S-5) whose corresponding eigenenergies are

$$N = 1 : \quad E_{D_\tau^\sigma} = \tau\Delta - \sigma g\mu_{\mathrm{B}}B/2 \qquad N = 2 : \quad \begin{cases} E_{S_0} = 3J/4 \\ E_{T_0^m} = -J/4 - mg\mu_{\mathrm{B}}B \\ E_{S_\tau} = 2\tau\Delta \end{cases} \tag{S-14}$$

For the moment we consider zero magnetic field $B = 0$ and can simplify the discussion by omitting the spin projection indices $\sigma$ in the doublets and $m$ in the triplet. In Sec. II C we will return to this notation when discussing the effect of a nonzero magnetic field. Clearly, in the $N = 1$ charge sector the ground state is the bonding doublet $D_-$ while for $N = 2$ the ground state is given by the localized singlet $S_-$ in the expected regime of weak exchange energy relative to the valley mixing, $J < 8\Delta$.

*b. Transport model* The source (s) and drain (d) leads are described as macroscopic reservoirs of noninteracting electrons through the Hamiltonian

$$H_{\mathrm{res}} = \sum_{rk\sigma}(\epsilon_{rk} + \mu_r)c_{rk\sigma}^\dagger c_{rk\sigma}, \tag{S-15}$$

where $c_{rk\sigma}^\dagger$ ($c_{rk\sigma}$) creates (annihilates) an electron in lead $r = \{\mathrm{s,d}\}$ with spin $\sigma = \{\uparrow,\downarrow\}$ and state index $k$. The eigenenergies of the leads are uniformly shifted by the bias voltage $V_{\mathrm{sd}}$ such that the electro-chemical potentials read $\mu_r = \pm V_{\mathrm{sd}}/2$ for $r = \{\mathrm{s,d}\}$, respectively. These reservoirs are assumed to be independently at equilibrium, characterized by a temperature $T$. We note that the matter of interactions in the CNT leads is a subtle one, but in the experiment we see no particular effect indicating their importance. Rather, the fact that we have quantized (yet broadened) states in the CNT leads seems to be important, resulting in an effective spin-dependence of the tunneling rates, see below.

The coupling between the dot and the leads is determined by the tunnel Hamiltonian

$$H_{\mathrm{tun}} = \sum_{rk\tau\sigma} t_{\tau\sigma}^r d_{\tau\sigma}^\dagger c_{rk\sigma} + \mathrm{H.c.}\,, \tag{S-16}$$

with the tunnel amplitudes $t_{\tau\sigma}^r$ assumed to be junction- ($r$), orbital- ($\tau$) and spin-dependent ($\sigma$). Since the tunnel rates required below have the form $2\pi\rho \left(t_{\tau\sigma}^r\right)^2$ (by Fermi's Golden Rule) these dependencies are modeled using three asymmetry parameters $\kappa$, $\gamma$ and $\zeta$, respectively:

$$t_{\tau\sigma}^r = \tfrac{1}{\sqrt{2}}(1 + \tau\kappa)(1 + \sigma\zeta)t^r, \qquad t^r = \frac{1 + r\gamma}{\sqrt{1 + \gamma^2}}\sqrt{\frac{(\Gamma/2)}{2\pi\rho}}, \tag{S-17}$$

where $r = \pm$ corresponds to $r = s/d$. Here $t^r$ characterizes the tunneling through each junction $r$ through the overall rates $\Gamma^r = 2\pi\rho(t^r)^2$ and $\rho$ is the density of states in the respective electrode. *We let $\Gamma = \Gamma^s + \Gamma^d$ characterize the overall scale of the rates which merely sets the magnitude of the current* and is irrelevant to the relative strengths of the different excitations which are of interest here. The latter are controlled by the quantum dot electron-vibrational states and the parameters $\gamma$, $\kappa$, and $\zeta$:

- In equation (S-17) we include the usual *junction asymmetry $\gamma = (t^s - t^d)/(t^s + t^d)$ for the tunnel coupling to the source and drain leads* through the last factor. Keeping $\Gamma = \Gamma^s + \Gamma^d$ fixed, a little algebra shows that $t^r$ is given by the second equation in equation (S-17). This asymmetry is relevant for modeling the measured differential conductance, which manifests some asymmetry between the intensities of positively and negatively sloped lines. We note that any $r$-dependence in the density of states in the leads that we ignored above can be absorbed into $\gamma$.

- The first factor in equation (S-17) captures an *orbital asymmetry $\kappa = (t_{+\sigma}^r - t_{-\sigma}^r)/(t_{+\sigma}^r + t_{-\sigma}^r)$* which is the same for all $\sigma$ and $r$. This derives from the linear combinations of the valley-dependent tunnel amplitudes, i.e. $t_{K\sigma}^r = (t_{+\sigma}^r + t_{-\sigma}^r)/\sqrt{2}$ and $t_{K'\sigma}^r = (t_{+\sigma}^r - t_{-\sigma}^r)/\sqrt{2}$. We notice that for strictly symmetric couplings of the two orbitals, $\kappa = t_{K'\sigma}^r/t_{K\sigma}^r = 1$, the tunnel Hamiltonian is symmetric with respect to the interchange of the $K$ and $K'$ valleys and cannot

induce transitions from the anti-symmetric $N = 1$ ground state $D_-$ into the symmetric state $N = 2$ ground state $S_-$. This would cause a strong suppression of low-bias transport up to a voltage where the lowest excitation for either $N = 1$ or $N = 2$ becomes accessible. This is not observed in the experiment and indicates that a definite orbital asymmetry is present, i.e., $\kappa \neq \pm 1$. (A similar problem arises for $\kappa = -1$.)

- Finally, we introduced in addition a possible *spin-dependence in the tunnel amplitudes* through the parameter $\zeta = (t^r_{\tau\uparrow} - t^r_{\tau\downarrow})/(t^r_{\tau\uparrow} + t^r_{\tau\downarrow})$ which is the same for all $\tau$ and $r$. Below we turn on $\zeta$ only in a nonzero applied magnetic field $B$ [cf. equation (S-59)]. As mentioned in the main article, this models a relevant aspect of the experiment, related to the fact that we have CNT leads, which we will discuss in Sec. II C when calculating the magnetic field evolution of the vibrational sideband lines observed in Fig. 3a of the main article.

From the above Hamiltonian we calculate the tunnel matrix elements (TMEs)

$$T^{r\sigma\eta}_{a\leftarrow b} = \sum_\tau t^r_{\tau\sigma} \langle a | d^\eta_{\tau\sigma} | b \rangle, \tag{S-18}$$

with the shorthand $d^\eta_{\tau\sigma} = d^\dagger_{\tau\sigma}, d_{\tau\sigma}$ for $\eta = \pm 1$. In Sec. II B we derive the explicit TMEs for the considered model after having considered the effect of the vibration, to which we turn now.

### 2. Coupling to the vibration - beyond the Anderson-Holstein model

The electronic model accounting for the many-electron states observed in the experiment is now extended to deal with the case where the coupling to the vibrational stretching mode of the nanotube is switched "on". As discussed in the main article, the differential conductance of Fig. 2d together with its magnetic field evolution in Fig. 3a shows several vibrational sidebands associated to the triplet state, but none for the ground singlet. This strong state-dependence in the coupling to the vibrational mode forces us to consider three different types of electron-vibration couplings which arise from the assumption of a linear dependence on the (dimensionless) mechanical displacement $Q$ of the nanotube in the electronic parameters $R = \{\varepsilon, \Delta, J\}$ of equation (S-5). In all cases, we assume $R(Q) = R + \lambda_R \hbar\omega\sqrt{2}Q$. We therefore have, in addition to the standard Holstein coupling $\lambda_\varepsilon$ to the number of particles $N$, a coupling $\lambda_\Delta$ which depends on the valley-mixing and a vibration-exchange coupling $\lambda_J$. By plugging these into equation (S-5) we arrive to the full Hamiltonian of the CNT quantum dot

$$H_{\text{qd}} = H_{\text{el}} + \frac{\hbar\omega}{2}\left[P^2 + (Q + \sqrt{2}\Lambda)^2\right] - \hbar\omega\Lambda^2, \tag{S-19}$$

which can be seen as a shifted quantum harmonic oscillator. The above electron-vibration couplings enter through the following operator

$$\Lambda = \lambda_\varepsilon N + \lambda_\Delta \sum_{\tau\sigma} \tau N_\tau - \lambda_J \mathbf{S}_+ \cdot \mathbf{S}_-, \tag{S-20}$$

which shifts the harmonic potentials associated to each electronic state (horizontal shift) and it also introduces a polaronic shift in the energy (vertical shift).

*a. Adiabatic potentials* To aid the intuition we consider the *adiabatic* potentials for this problem, obtained by treating $Q$ as a classical variable. These potentials were plotted in the main article in Fig. 4. The adiabatic potentials associated to the electronic states $|e\rangle$, where $e = D_\pm, S_\pm, S_0, T_0$, can be characterized by polaron shift of its potential minimum $\Lambda_e = \langle e | \Lambda | e \rangle$, i.e.

$$\Lambda_{D_\tau} = \lambda_\varepsilon + \tau\lambda_\Delta, \qquad\qquad \Lambda_{S_\tau} = 2\lambda_\varepsilon + 2\tau\lambda_\Delta, \tag{S-21}$$

$$\Lambda_{S_0} = 2\lambda_\varepsilon + \frac{3}{4}\lambda_J, \qquad\qquad \Lambda_{T_0} = 2\lambda_\varepsilon - \frac{1}{4}\lambda_J. \tag{S-22}$$

Roughly speaking, the magnitude of these shifts determine whether none, several or many vibrational sidebands will appear (see below). These expressions reveal that the polaronic shifts corresponding to the various electronic states can indeed be different once one abandons the

simplifying assumption of the Anderson-Holstein model, that the *electronic excitations* are not affected by a distortion (the vibration coordinate $Q$):

(i) Both the valley-vibration coupling $\lambda_\Delta$ and the exchange-vibration coupling $\lambda_J$ distinguish multiplets with zero $(S_0, T_0)$ and maximal orbital polarization $(D_\tau, S_\tau)$.

(ii) Also, the exchange-vibration coupling allows for a fine tuning of the coupling to the vibration for the $S_0$ and $T_0$ states. As we will show in Sec. II C, the horizontal shifts of the ground singlet $S_-$ and the triplet $T_0$ can be tuned so they are qualitatively different, giving rise to contrasting Franck-Condon factors.

Although the intuition is useful, in our calculations we account for the exact many-body eigenstates of the quantum-dot Hamiltonian in equation (S-19). These are obtained as a tensor product of the electronic states $|e\rangle$ and the quantum vibrational states $|\nu\rangle$. These electron-vibration eigenstates $|e, \nu\rangle$ thus yield the following eigenenergies for $e = D_\pm, S_\pm, S_0, T_0$:

$$E_{e,\nu} = \langle e, \nu | H_{\text{qd}} | e, \nu \rangle = E_e - \hbar\omega \Lambda_e^2 + \frac{\hbar\omega}{2}\left(\nu + \frac{1}{2}\right). \tag{S-23}$$

The horizontal shift in equation (S-19), although not present in the above energies, enters as the equilibrium position for the vibrational states, and thereby it plays a crucial role in the transition amplitudes when calculating the matrix elements of the tunnel Hamiltonian, as we will show in Sec. II B.

Finally, we note that in the limit $\lambda_\Delta = \lambda_J = 0$ the model reduces the standard Anderson-Holstein model. Here, the polaronic energy shifts can be absorbed into the gate voltage since the electron-vibration coupling for transitions involving states differing by one electron is, in all cases, the same.

*b. State-dependent Franck-Condon shifts - the role of $\lambda_\Delta$ and $\lambda_J$* The tunnel amplitudes and thus the intensities of the lines in the differential conductance depend on the Franck-Condon amplitudes that can be calculated from the relative shifts between the harmonic potentials, i.e. $\Lambda_{x,y} = \Lambda_x - \Lambda_y$ where $x, y$ are labels of the many-particle states differing by one electron. Of particular relevance to the experiment are the two relative oscillator shifts

$$\Lambda_{S_-, D_-} = \lambda_\varepsilon - \lambda_\Delta, \tag{S-24}$$

$$\Lambda_{T_0, D_-} = \lambda_\varepsilon + \lambda_\Delta - \frac{1}{4}\lambda_J, \tag{S-25}$$

for transitions from the ground doublet $D_-$ to the ground singlet $S_-$ and to the triplet $T_0$, respectively. Since $\lambda_\Delta$ is present in both transitions but with different sign, it can induce different coupling strengths for the singlet and the triplet. Additionally, $\lambda_J$ can be used to tune the coupling strength of the triplet independently of the singlet. According to equations (S-57)-(S-23), the couplings $\lambda_\Delta$ and $\lambda_J$ also induce vertical polaronic shifts (i.e., of the energy at the minimum) which are given by

$$\Lambda_{S_-}^2 = 4(\lambda_\varepsilon - \lambda_\Delta)^2, \tag{S-26}$$

$$\Lambda_{T_0}^2 = (2\lambda_\varepsilon - \frac{1}{4}\lambda_J)^2. \tag{S-27}$$

Here it is important to note that the triplet polaronic shift does not depend on $\lambda_\Delta$, and in consequence its coupling to the vibration can be modified through $\lambda_\Delta$ in equation (S-25), whereas the polaronic shift in equation (S-27) remains unaffected. The coupling $\lambda_J$ can therefore be used to tune the polaronic shift to the triplet without affecting the singlet $S_-$ state. It is thus indeed possible with our model to describe the central observation in the experiment.

### 3. Spin-orbit interaction effects

In this section we consider the possible effects on the electronic properties of the many-body states of the CNT quantum dot due to a static curvature-enhanced spin-orbit (SO) coupling $\Delta_{\text{SO}}$. There are two reasons for this: First, one may wonder how the above scenario is affected by $\Delta_{\text{SO}}$ in general. Second, even when $\Delta_{\text{SO}}$ is too small as to produce a significant shift in the line positions of the differential conductance, the SO mixing turns out to be crucial to give a nonvanishing amplitude to the ground singlet line in Fig. 3c.

When considering the nanotube axis oriented along the $x$-direction as we do here, this SO interaction enters through the following Hamiltonian[6]

$$H_{\mathrm{SO}} = \Delta_{\mathrm{SO}} \sigma_x \otimes \tau_z \,, \tag{S-28}$$

where $\sigma_x$ and $\tau_z$ are Pauli matrices in spin and valley subspaces, respectively. Hence, the spin-orbit interaction conserves the valley structure, but it couples different spins, yielding spin-flip processes. Written in the BA-basis,

$$H_{\mathrm{SO}} = \Delta_{\mathrm{SO}} \sum_{\tau\sigma} d_{\tau\sigma}^{\dagger} d_{\bar{\tau}\bar{\sigma}} \,, \tag{S-29}$$

the spin-orbit term couples the Kramers doublets $|D_-^{\uparrow}\rangle \leftrightarrow |D_+^{\downarrow}\rangle$ and $|D_-^{\downarrow}\rangle \leftrightarrow |D_+^{\uparrow}\rangle$. Here we used the compact notation $\bar{\sigma} = -\sigma$ and $\bar{\tau} = -\tau$.

Naively, one expects that for $\Delta_{\mathrm{SO}} < \hbar\omega\Lambda_{T_0 S_-}$ the state-dependent vibrational coupling will survive. To again develop some better intuition we discuss as before the adiabatic approximation where the mechanical displacement $Q$ is considered as classical parameter. In the full numerical calculations, however, we always exactly diagonalize the quantum-dot Hamiltonian model.

*a. One-electron states* The adiabatic Hamiltonian matrix in the $N = 1$ sector contains a SO mixing term between the doublets:

$$H_{\mathrm{ad}}^{(1)} = \begin{pmatrix} E_{D_-^{\uparrow}}(Q) & 0 & 0 & \Delta_{\mathrm{SO}} \\ 0 & E_{D_-^{\downarrow}}(Q) & \Delta_{\mathrm{SO}} & 0 \\ 0 & \Delta_{\mathrm{SO}} & E_{D_+^{\uparrow}}(Q) & 0 \\ \Delta_{\mathrm{SO}} & 0 & 0 & E_{D_+^{\downarrow}}(Q) \end{pmatrix} \,, \tag{S-30}$$

where the electronic states are now associated to the following adiabatic potentials

$$E_{\mathsf{x}}(Q) = E_{\mathsf{x}} - \hbar\omega\Lambda_{\mathsf{x}}^2 + \frac{\hbar\omega}{2}(Q + \sqrt{2}\Lambda_{\mathsf{x}})^2 \,. \tag{S-31}$$

We therefore expect the SO interaction to couple the doublets at the points of intersection of the adiabatic potentials as $Q$ is varied. For $\Delta_{\mathrm{SO}} < \Delta$, the minimum of the ground doublet parabola is conserved, and one might expect little change for the line positions in the differential conductance. In Fig. S-8a we show the one-electron adiabatic potentials including a small spin-orbit coupling $\Delta_{\mathrm{SO}} = 0.125\Delta = 0.1$ meV and compare them to the bare adiabatic potentials $E_{D_-}$. The important conclusion to draw here is that in the parameter regime that we will discuss below in Sec. II C there is no significant change around the ground state potential energy minimum.

*b. Two-electron states - state dependent Franck-Condon shifts* We now discuss the situation in the two-electron charge state. The form of the SO Hamilonian in equation (S-29) shows that this operator flips both the orbital and spin projection in one of the two electrons in the quantum dot. Therefore, the two states $T_0^0$ and $S_0$ with electrons with opposite orbitals are not affected by this coupling due to Pauli's exclusion principle. The other states $S_\tau$ and $T_0^{\pm 1}$ are indeed mixed, as revealed by the Hamiltonian matrix block:

$$H_{\mathrm{ad}}^{(2)} = \begin{pmatrix} E_{S_-}(Q) & \Delta_{\mathrm{SO}} & 0 & -\Delta_{\mathrm{SO}} & 0 & 0 \\ \Delta_{\mathrm{SO}} & E_{T_0^1}(Q) & 0 & 0 & 0 & -\Delta_{\mathrm{SO}} \\ 0 & 0 & E_{T_0^0}(Q) & 0 & 0 & 0 \\ -\Delta_{\mathrm{SO}} & 0 & 0 & E_{T_0^{-1}}(Q) & 0 & \Delta_{\mathrm{SO}} \\ 0 & 0 & 0 & 0 & E_{S_0}(Q) & 0 \\ 0 & -\Delta_{\mathrm{SO}} & 0 & \Delta_{\mathrm{SO}} & 0 & E_{S_+}(Q) \end{pmatrix} \,. \tag{S-32}$$

Again, we expect the spin-orbit coupling to be only important around the crossings of the adiabatic potentials as $Q$ is varied. The above matrix thus mixes the two-particle states as follows: (i) The localized singlets $S_\tau$ will be repelled by the triplets $T_0^{\pm 1}$. (ii) These triplet states slightly repel each other due to a second order process (since there is no direct matrix element connecting them).

It now depends on the strength of the spin-orbit coupling relative to the vibrational coupling energy (shifts of the potential energy minima $\times \hbar\omega$) and the splitting between the "bare" electronic states (vertical energies at the minima) whether the SO coupling has a negligible impact on our

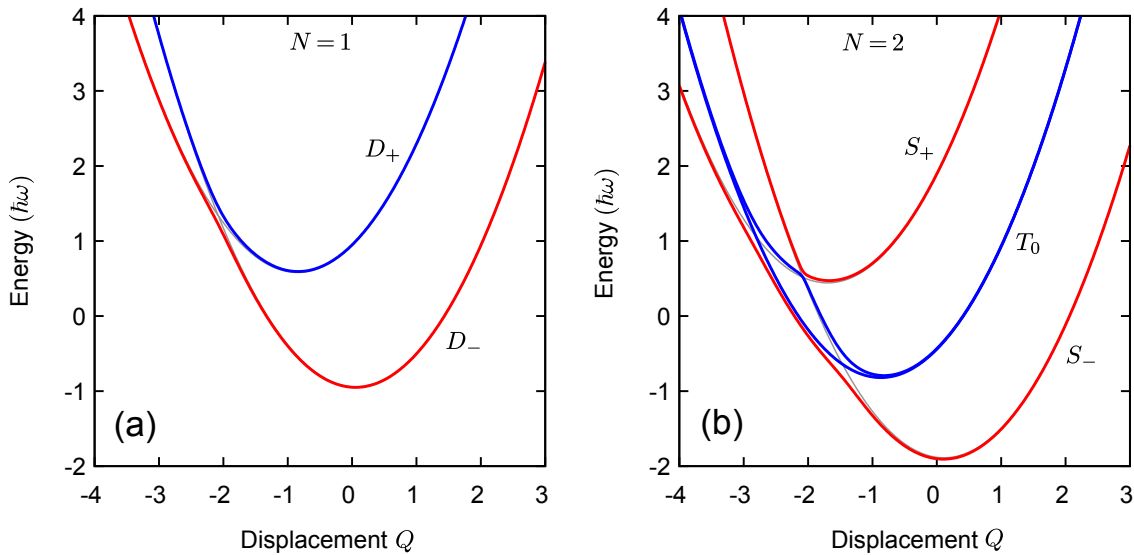

FIG. S-8. Adiabatic potentials including spin-orbit interaction $\Delta_{SO} = 0.125\Delta = 0.1$ meV. (a) One-particle mixed doublets from equation (S-30). (b) Two-particle potentials calculated from equation (S-32) in the subspace mixed by the SO coupling. Adiabatic potentials for $\Delta_{SO} = 0$ are shown in grey for comparison. For large $Q$ we can associate each adiabatic potential to the marked electronic states. In both charge sectors, the relevant lower minima are slightly affected by $\Delta_{SO}$.

mechanism or not: the SO coupling only has a big effect at crossings of potential energies which may be far away from the relevant minima. For the estimated experimental parameters, the energy difference between the triplet and the ground singlet is noticeable. Moreover, the adiabatic potentials related to these states are strongly shifted, meaning that the SO coupling will not have a strong impact on the development of the minima of those if $\Delta_{SO} < \hbar\omega\Lambda_{T_0,S_-}$. This is corroborated by Fig. S-8b where the two-electron adiabatic potentials are shown. This is in agreement with the measurements. However, even in this case the SO coupling plays a role: it is required for the explanation of the peak amplitudes in a magnetic field as we discuss in Sec. II C 2 b.

For strong spin-orbit coupling, $\Delta_{SO} \geq \hbar\omega\Lambda_{T_0,S_-}$, the situation is quite different: All involved Franck-Condon shifts of the vibrational mode become approximately the same, in clear disagreement with the measurements. Roughly speaking, for strong spin-orbit coupling the states $S_-$ and $T_0$ are strongly mixed and the difference in their coupling to the vibration is "averaged out".

We conclude two things: (i) The SO interaction (S-28)-(S-29) *does not generate a state-dependent electron-vibration coupling.* (ii) When present and strong, the SO interaction *rather tends to weaken it*, merely renormalizing the vibration frequency.

### B. Master equations - tunneling and relaxation

In this section we describe the employed method for the calculation of the Coulomb-diamond stability diagrams shown in Figures 2f - 2g and Fig. 3c of the main article.

#### 1. Tunnel processes

In the stationary limit and for weak couplings to the source and drain leads, the occupations probabilities $p_a$ in the dot obey the rate equations

$$0 = \sum_{b \neq a} (W_{ab} p_b - W_{ba} p_a), \tag{S-33}$$

where $W_{ab} = W_{ab}^s + W_{ab}^d$ represents the probability per unit time for a state transition $|b\rangle \to |a\rangle$ in the quantum dot. Since we will restrict ourselves to lowest order contributions in $\Gamma$, the overall

scale of $W_{ab}$, the above rates coincide with those obtained by Fermi's Golden's Rule, namely

$$W_{ab} = \sum_r W_{ab}^r = \sum_{r\eta} \Gamma_{ab}^{r\eta} f_r^{\eta}(E_a - E_b)\,, \tag{S-34}$$

where $f_r^{\eta}(x) = [1 + \exp(\eta(x - \mu_r)/k_{\mathrm{B}}T)]^{-1}$ is the Fermi distribution function for an electron ($\eta = +1$) or a hole ($\eta = -1$) and the sum runs over the reservoirs $r = \{\mathrm{s,d}\}$. The tunnel current $I_r$ that flows out of electrode $r$ is calculated through the standard master equation approach in the single-electron tunneling regime (SET)

$$I_r = \sum_{ab} (N_a - N_b) W_{ab}^r p_b\,, \tag{S-35}$$

where $N_a = \langle a|N|a\rangle$ is the electron number in the quantum-dot state $|a\rangle$. Like the state occupation probabilities $p_a$ obtained from equation (S-33), the current thus also depends on the tunneling rates $\Gamma_{ab}^{r\eta}$. The tunnel rates $\Gamma_{ab}^{r\eta}$ are related to the tunnel matrix elements (TMEs) $T_{a\leftarrow b}^{r\sigma\eta}$ for an electron with spin $\sigma$ entering ($\eta = +1$) or leaving ($\eta = -1$) the dot and the density of states in the leads $\rho$. The latter is assumed to be constant in the model (wide-band limit) and hence

$$\Gamma_{ab}^{r\eta} = 2\pi \sum_{\sigma} |T_{a\leftarrow b}^{r\sigma\eta}|^2 \rho\,. \tag{S-36}$$

In equation (S-18) we only accounted for the tunnel matrix elements associated to the pure electronic states $|e\rangle$, i.e. the many-body eigenstates of equation (S-5) labeled by $e = D_{\pm}, S_{\pm}, S_0, T_0$. These now need to be extended to the electron-vibration states $|e, \nu\rangle$ by adding the Frank-Condon overlap $F_{\nu',\nu}$ of the vibrational wave-functions involved in the tunnel event, i.e.

$$T_{a,\nu'\leftarrow b,\nu}^{r\sigma\eta} = F_{\nu',\nu} T_{a\leftarrow b}^{r\sigma\eta}\,. \tag{S-37}$$

where now $a, b = D_{\pm}, S_{\pm}, S_0, T_0$ and $T_{a\leftarrow b}^{r\sigma\eta}$ is given by equation (S-18). The Franck-Condon coefficient strongly depends on the horizontal shift $\lambda$ between the adiabatic potentials of the electronic states. If the dot is initially in the state $|b, \nu\rangle$, the probability of a transition to a final state $|a, \nu'\rangle$ shifted in $\lambda$ is modulated by[7–10]

$$F_{\nu',\nu} = \langle \nu'|e^{i\sqrt{2}\lambda P}|\nu\rangle = e^{-\lambda^2/2}(-\lambda)^{\nu'-\nu}\sqrt{\frac{\nu!}{\nu'!}}L_{\nu}^{\nu'-\nu}(\lambda^2)\,, \tag{S-38}$$

for $\nu' \geq \nu$ (replace $\nu' \leftrightarrow \nu$ for $\nu \geq \nu'$) and $L_j^i(x)$ is the associated Laguerre polynomial. Since the $\lambda$-shift strongly attenuates the transition probability between the involved electronic states at different regimes of the bias, one of the key aspects of the model is that *electronic transitions to the triplet state allow a change in the number of vibrational quanta while this is exponentially suppressed for transitions to the ground singlet*. We note that also in the presence of vibrations *the overall tunneling rate $\Gamma$* – entering as an overall factor through equation (S-17), equation (S-18) and equation (S-37) – *merely sets the scale of the current* and is irrelevant to the relative magnitude of the different excitations which is of interest here.

We now proceed with the explicit calculation of the TMEs for the $1 \leftrightarrow 2$ transitions. Let us consider "charging" transitions, i.e., those that start from the $N = 1$ charge sector [the "discharging" transitions involve the TMEs $T_{b\leftarrow a}^{r\sigma\eta} = (T_{a\leftarrow b}^{r\sigma\eta})^*$]. For a strong intrinsic relaxation (see below), the relevant transitions are those which begin from $|D_-, 0\rangle$, i.e., the ground doublet $D_-$ and no vibration, $\nu = 0$. For (charging) transitions to a doubly occupied state we use $\eta = 1$ and find the following amplitudes

$$T_{S_{\tau,\nu}\leftarrow D_{-,0}^{r\sigma'}} = F_{\nu,0}(\Lambda_{S_{\tau},D_-})\bar{\sigma}t_{-\sigma'}^r \delta_{\tau-\bar{\sigma}}\delta_{\sigma'\bar{\sigma}}\,, \qquad T_{T_{0,\nu}\leftarrow D_{-,0}^{r\sigma'}} = F_{\nu,0}(\Lambda_{T_0,D_-})\bar{\tau}t_{+\sigma'}^r \delta_{\sigma\uparrow}\delta_{\sigma'\uparrow}\,, \tag{S-39}$$

$$T_{T_{0,\nu}\leftarrow D_{-,0}^{r\sigma'}} = F_{\nu,0}(\Lambda_{T_0,D_-})\bar{\tau}\frac{t_{+\sigma'}^r}{\sqrt{2}}\delta_{\sigma'\bar{\sigma}}\,, \qquad T_{T_{0,\nu}^{-1}\leftarrow D_{-,0}^{r\sigma'}} = F_{\nu,0}(\Lambda_{T_0,D_-})\bar{\tau}t_{+\sigma'}^r \delta_{\sigma\downarrow}\delta_{\sigma'\downarrow}\,, \tag{S-40}$$

$$T_{S_{0,\nu}\leftarrow D_{-,0}^{r\sigma'}} = F_{\nu,0}(\Lambda_{S_0,D_-})\bar{\sigma}\frac{t_{+\sigma'}^r}{\sqrt{2}}\delta_{\sigma'\bar{\sigma}}\,. \tag{S-41}$$

Here we again used the compact notation $\bar{\sigma} = -\sigma$ and $\bar{\tau} = -\tau$. We note that the transition to the $S_+$ singlet from the ground doublet is forbidden in the SET regime since it would involve an orbital

flip process which is not present in the tunnel Hamiltonian. The same restriction also applies to the $S_- \leftrightarrow D_+$ transitions. However, as discussed in the previous section, a small spin-orbit term hybridizes the doublets and localized singlet states, making visible the resonance line associated to this last transition (line E in Fig. 1d of the main article).

### 2. Intrinsic relaxation

In the measured stability diagrams of the main article, we note that electronic transitions starting from an excited state are strongly suppressed. In order to account for this effect in the most simple way we model the influence of an environmental bath by allowing for relaxation processes. Since the main source of excitations is the electron transport we neglect absorption processes due to the bath by setting the bath temperature $T_b = 0$. Furthermore, we allow in each charge sector for the relaxation of any energetically higher-lying state $b$ into any lower-lying state $a$, independently of the spin or the orbital distribution of the involved states. *The decay rates are assumed to be proportional to the energy difference between these states and is assumed to exceed the tunneling relaxation rates:*

$$W_{ab}^{\mathrm{rel}} = \Gamma_{\mathrm{rel}}(E_b - E_a), \qquad \Gamma_{\mathrm{rel}}(E) = \Gamma \times (E/0.2 \ \mathrm{meV}). \tag{S-42}$$

The relaxation rate matrix $W^{\mathrm{rel}}$ is thus upper triangular and it is specified in units of $\Gamma$. The corresponding rates $W_{ab}^{\mathrm{rel}}$ are added to the golden rule rates of equation (S-34) and the master equations are solved using these modified rates. This depends little on the details and has the main effect of preventing a very strong nonequilibrium state on the quantum dot, enhancing the ground state occupation probability in each charge sector.

### C. Comparison with experiment

In this section we describe how we proceed in finding a unique parameter *regime* of the model which is able to qualitatively explain the experimental data. The key effect of state-dependent vibrational coupling is discussed fully in the zero magnetic field case. The finite magnetic field experiment brings in some complications due to spin-dependent tunneling and spin-orbit coupling, which are, however, not crucial for the central point of the article. Finally, we discuss the dependence on the parameters and identify the problem that a standard Anderson-Holstein model has with explaining the observations.

#### 1. Zero magnetic field

*a. Vibrations "off"* As already outlined in the experimental Sec. I B we calculate the capacitive effect induced by the gates and the source and drain leads from Fig. S-9a. The left and right resonance lines (red dashed lines in Fig. S-9a), related to the transition between the $N = 1$ and $N = 2$ ground states $D_- \leftrightarrow S_-$, are described by the following relations between the top-gate and the source-drain voltages

$$V_{\mathrm{sd}}^{\mathrm{s}} = -\frac{2\alpha_{\mathrm{tg}}}{1 + (\alpha_{\mathrm{s}} - \alpha_{\mathrm{d}})} V_{\mathrm{tg}} = m_{\mathrm{s}} V_{\mathrm{tg}}, \tag{S-43}$$

$$V_{\mathrm{sd}}^{\mathrm{d}} = +\frac{2\alpha_{\mathrm{tg}}}{1 - (\alpha_{\mathrm{s}} - \alpha_{\mathrm{d}})} V_{\mathrm{tg}} = m_{\mathrm{d}} V_{\mathrm{tg}}. \tag{S-44}$$

By extracting the slopes of these lines ($m_{\mathrm{s}} = -1.1$ and $m_{\mathrm{d}} = 0.9$) we obtain the following values for the lever arms: $\alpha_{\mathrm{tg}} = 0.5$ and $\alpha_{\mathrm{s}} - \alpha_{\mathrm{d}} = -0.1$. We next extract the electronic parameters, the subband splitting $\Delta$ and the exchange energy $J$. To this end, we use the formulas [cf. equation (S-14), (S-45), and (S-46) below] for the bias distances $\mu_1 = 1.2$ meV and $\mu_2 = 2.7$ meV in energy units ($|e| = 1$), measured in Fig. S-9a:

$$\mu_1 = E_{T_0} - E_{S_-} = 2\Delta - J/4, \tag{S-45}$$

$$\mu_2 = E_{S_0} - E_{S_-} = 2\Delta + 3J/4, \tag{S-46}$$

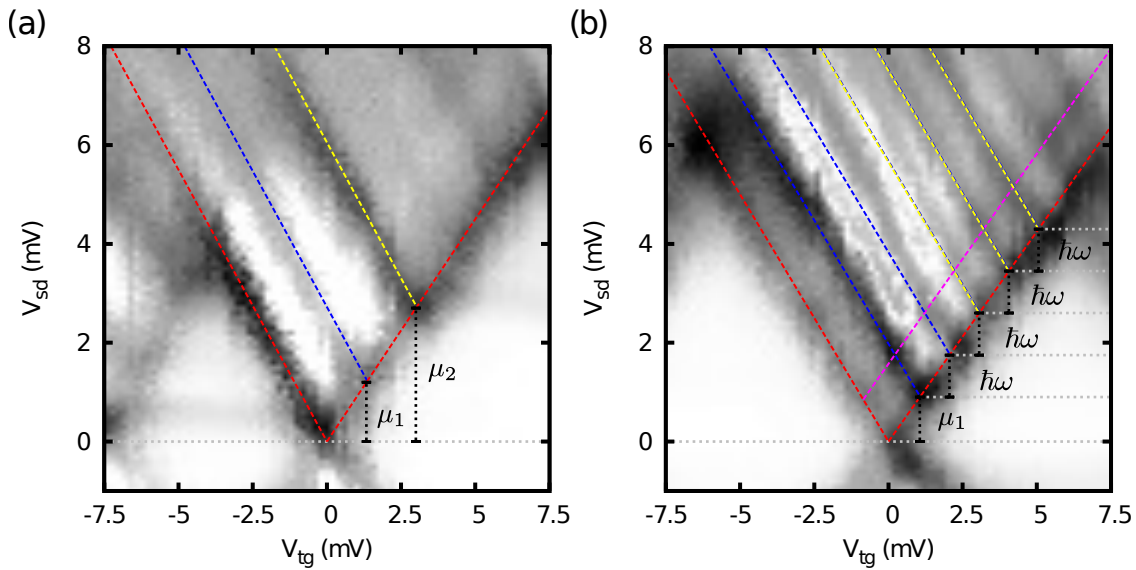

FIG. S-9. Resonance lines fitting for $N = 1 \leftrightarrow 2$ transitions. (a) Vibrations switched "off" [zoom-in of Fig. S-6b and Fig. 2c in the main article]: The measured $\mu_1$ and $\mu_2$ voltages are related, respectively, to electronic transition energies $E_{T_0} - E_{S_-}$ and $E_{S_0} - E_{S_-}$. (b) Vibrations switched "on" [zoom-in of Fig. S-6d and Fig. 2d in the main article]: The measured $\mu_1$ here corresponds to $E_{T_0} - E_{S_-}$. In both panels, dashed lines are visible transitions described by the model: $|D_-, 0\rangle \leftrightarrow |S_-, 0\rangle$ (red), $|D_-, 0\rangle \rightarrow |T_0, \nu\rangle$ with $\nu = 0 - 4$ (blue), $|D_-, 0\rangle \rightarrow |S_0, \nu\rangle$ with $\nu = 0 - 2$ (yellow). Yellow lines are on top of blue lines (see condition 3 in the text). The $|S_-, 0\rangle \rightarrow |D_-, 1\rangle$ transition is shown in panel (b) in dashed magenta and is just discernible in the measured differential conductance.

We obtain $\Delta = (\mu_2 + 3\mu_1)/8 = 0.8$ meV and $J = \mu_2 - \mu_1 = 1.5$ meV [cf. equation (S-3)].

    *b. Vibrations "on"* In the regime where the CNT quantum dot is coupled to the vibrational mode, we aim to account for the following features observed in the experimental data of Fig. S-9b (Fig. 2d of the main article):

1. The experiment shows that the two-electron ground state is given by the singlet $S_-$ and the first excited state is the triplet $T_0$. In addition, their line positions remain almost unchanged when turning "on" the vibrational coupling. In our model we thus include the base values of $J$ and $\Delta$, fixed by their fitting to the electronic spectrum in Fig. S-9a. When allow these energies to become sensitive to the vibration $Q$ through $\lambda_\Delta$ and $\lambda_J$ we must thus impose the severe restriction that these couplings do not give rise to a polaronic shift.

2. The ground singlet $S_-$ transition is very weakly coupled to the vibrational mode, the only visible lines of the $|D_-, \nu\rangle \leftrightarrow |S_-, \nu'\rangle$ transitions is the "zero-phonon" one, $\nu = \nu' = 0$. We thus require a small value for the horizontal shift between the $N = 1$ and $N = 2$ ground state potential minima, i.e. $|\Lambda_{S_-, D_-}| \ll 1$. In contrast, the triplet transition shows a whole series of vibrational sidebands and must therefore require a sizeable coupling, i.e., $|\Lambda_{T_0, D_-}| \simeq 0.5$, cf. Fig. 4 of the main article.

3. Although with vibrational coupling turned "off" the excited singlet $S_0$ transition appears in Fig. S-9a, no transitions $|D_-, 0\rangle \leftrightarrow |S_0, \nu\rangle$ for $\nu = 1, 2, 3, \ldots$ can be distinguished in the stability diagram Fig. S-9b when they are turned "on". The approximate relation $J \approx 2\hbar\omega$ suggests these transitions are in fact superimposed with the triplet sidebands: the energy difference between $S_0$ and $T_0$, given by the exchange energy $J$, happens to be commensurate with the vibration energy $\hbar\omega$ to within the experimental thermal line broadenings. We require only $J = n\hbar\omega$ and determine the best fitting integer $n$ (confirming indeed that $n = 2$).

In applying these constraints to the model parameters, we used energies obtained from lever arms that we extracted independently for Fig. S-9b: in this case we obtain from the slopes ($m_s = -1.0$,

$m_d = 0.85$) of the ground state lines, giving the following values: $\alpha_{tg} = 0.46$ and $\alpha_s - \alpha_d = -0.08$. For the vibration energy we use the $\hbar\omega = 0.85$ meV, the fitted value of the mean level spacing $\Delta E_{vib}$, obtained specifically for the upper panel in Fig. 2h of the main article.

According to constraint (1) for fixed electronic parameters $\Delta$ and $J$, we need to adjust the energy difference between $E_{T_0,0}$ and $E_{S_-,0}$ to the measured bias $\mu_1 \simeq 0.9$ mV in Fig. S-9b. This implies

$$\mu_1 = E_{T_0,0} - E_{S_-,0} = 2\Delta - \frac{J}{4} + \hbar\omega(\Lambda_{S_-}^2 - \Lambda_{T_0}^2).$$ (S-47)

For the following algebra it is convenient to denote the two free parameters by $x = \Lambda_{S_-}$ and $y = \Lambda_{T_0}$ and introduce $p = (\mu_1 - 2\Delta + J/4)/\hbar\omega \simeq -0.382$. The first condition for the couplings then reads:

$$y = \sqrt{x^2 - p}.$$ (S-48)

The horizontal shift $q = \Lambda_{T_0,D_-}$ between the triplet and the doublet can be written as $q = y - x/2$ [cf. equation (S-21)-(S-22)] and hence we have

$$y = \frac{x}{2} + q.$$ (S-49)

From condition 2 we have $|x| \ll 1$ and $|q| \sim 1$. Bearing in mind this restriction, we find a unique solution for $x$ when requiring both the above two equations to hold simultaneously:

$$x = \frac{2q}{3} \pm \frac{4}{3}\sqrt{q^2 + \frac{3}{4}p}.$$ (S-50)

Notice here that we cannot take $|q| < \sqrt{3|p|/4} \simeq 0.54$ since this would imply no real-valued solution. On the other hand, we cannot increase $|q|$ indefinitely either since otherwise the solution for $x$ grows and will violate condition 2 [i.e., the sidebands associated to $S_-$ would become visible]. For $q$ we pick the value $q = 0.6$ which gives the best overall agreement with the experiment (number of visible triplet sidebands), implying $x = -0.04$ and $y = 0.62$.

Finally, we use the condition 3 to obtain the three vibration couplings. This last requirement reads with $z = \Lambda_{S_0}$ and equation (S-47)

$$n\hbar\omega \sim E_{S_0,0} - E_{T_0,0} = J - \hbar\omega(z^2 - y^2),$$ (S-51)

where $n$ is an integer number. By using $J = 1.5$ meV and $\hbar\omega = 0.85$ meV, we obtain $y^2 + J/\hbar\omega \simeq 2.15$ and this implies that $n$ cannot be larger than 2. As mentioned above, $n = 2$ yields approximately the same line position for $S_0$ as observed before in the pure electronic regime. Solving the above equation with $n = 2$ for $z$ and using the polaronic shifts definitions in terms of the $\lambda$-parameters (equations (S-21) and (S-22)), we obtain the parameter values used in Figures 2d and 3c in the main article,

$$\lambda_\varepsilon = 0.28, \quad \lambda_\Delta = 0.32, \quad \lambda_J = -0.22.$$ (S-52)

We emphasize that the above procedure essentially determines a *unique regime* of parameters consistent with the experimental results. This concerns their qualitative features: conditions 2 and 3 only define a *range* of possible values for the polaronic shifts and therefore small deviations of the above obtained values for the $\lambda$-parameters produce similar results. However, *an Anderson-Holstein type model ($\lambda_\Delta = \lambda_J = 0$) is certainly not consistent with the measurements, see also Sec. II C 3. Thus, despite the fact that there are several parameters, the experiment imposes strong restrictions, in particular, limiting the choice of vibrational couplings. Importantly, the transport parameters $\gamma$, $\kappa$ and $\zeta$ adjust other aspects of the transport spectrum but do not generate or affect in an essential way the state-dependent vibrational coupling.* In particular at $B = 0$ are not that important.

### 2. Magnetic field spectroscopy

We now consider the predictions of the above described model for magnetic field evolution of the $dI/dV_{sd}$-peak intensities shown in Fig. 3 of the main article.

*a. Zeeman effect on state-dependent vibrational side bands*    When applying a static magnetic field $B$, the many-electron states in the quantum dot experience a Zeeman shift. Since the field is perpendicular to the CNT axis the orbital splitting due to $B$ can be neglected[11]. The field dependence of the doublet and triplet states with spin projection indices $\sigma = \{+1, -1\}$ and $m = \{+1, 0, -1\}$, respectively, is

$$E_{D_\pm^\sigma, \nu} = E_{D_\pm, \nu}^{(0)} - \tfrac{1}{2}\sigma g\mu_\mathrm{B} B \,, \tag{S-53}$$

$$E_{T_0^m, \nu} = E_{T_0, \nu}^{(0)} - mg\mu_\mathrm{B} B \,, \tag{S-54}$$

where $E_{x,\nu}^{(0)}$ denote the zero-field eigenenergies (S-23). The singlets energies are obviously independent of $B$. (Regarding the negligble importance of the singlet $S_+$, see the remark in the main article's text and the footnote 28 there.) Since the $N = 1$ quantum dot ground state ($|D_-^\uparrow, 0\rangle$, $\nu = 0$) evolves with $-g\mu_\mathrm{B} B/2$, the energy cost for adding one electron to the quantum dot depends in all cases on the magnetic field strength, no matter which final state is reached: relative to this state, the addition energies for the $N = 2$ states accessible in the sequential tunneling regime (see Sec. II B) are

$$\Delta E_{S_-, \nu} = \Delta E_{S_-, \nu}^{(0)} + \tfrac{1}{2}g\mu_\mathrm{B} B \,, \tag{S-55}$$

$$\Delta E_{T_0^1, \nu} = \Delta E_{T_0, \nu}^{(0)} - \tfrac{1}{2}g\mu_\mathrm{B} B \,, \tag{S-56}$$

$$\Delta E_{T_0^0, \nu} = \Delta E_{T_0, \nu}^{(0)} + \tfrac{1}{2}g\mu_\mathrm{B} B \,, \tag{S-57}$$

$$\Delta E_{S_0, \nu} = \Delta E_{S_0, \nu}^{(0)} + \tfrac{1}{2}g\mu_\mathrm{B} B \,. \tag{S-58}$$

where $\Delta E_{x,\nu}^{(0)} = E_{x,\nu}^{(0)} - E_{D,0}^{(0)}$ are the zero-field values. The ground-state singlet $S_-$ and the excited triplet $T_0^1$ lines thus evolve with opposite slopes as function of $B$. Since the triplet $T_0$ states have the interesting vibrational side bands – both in the experiment and in our calculations – their field dependence is plotted such that the triplet peak positions stay fixed, both here in Fig. S-10 and in Figures 3a - 3c of the main article. In the theoretical plots this is achieved by replacing $V_\mathrm{tg} \to V_\mathrm{tg} + \tfrac{1}{2}g\mu_\mathrm{B} B$.

In Fig. S-10a we show the magnetic-field evolution of the calculated differential conductance along the line $V_\mathrm{sd} = V_\mathrm{tg} + 5$ mV for spin-independent tunnel barriers, i.e., $\zeta = 0$ for all $B$ (but including SO coupling). What the model accounts for at this level relates to the key observation made in the measurements: the triplet $T_0^+$ state (vertical) show vibrational side bands, whereas the *ground* singlet state $S_-$ (left most sloped line) does not. Importantly, the other sloped lines should *not* be confused with vibrational side bands of the ground singlet $S_-$: These are the transitions into the $m = 0$ component of the triplet, $T_0^0$, or the *excited* singlet $S_0$ (the latter is shifted in about $2\hbar\omega$ with respect to the former, see above in Sec. II C 1) and their vibrational side bands. That these lines should be there is clear from the theoretical model and its analysis [cf. equation (S-57) and equation (S-58)] but is also borne out by noting the difference between the distance A and the following ones B-D [just as in the experimental data in Fig. 3a of the main article]. In the experiment these Zeeman split-off triplet states in Fig. S-10a, as well as the singlet $S_0$ are not observed in Fig. 3a of the main article.

*b. Suppression of Zeeman split-off states*    In order to rationalize the observed absence of Zeeman split-off states in the experiment we now include a spin-dependence in the tunnel amplitudes through the parameter $\zeta$ [cf. equation (S-17)] which we take to be $B$-dependent:

$$\zeta(B) = \frac{1 - e^{-g\mu_\mathrm{B} B / k_\mathrm{B} T}}{1 + e^{-g\mu_\mathrm{B} B / k_\mathrm{B} T}} \,, \tag{S-59}$$

This phenomenological function ensures that for $g\mu_\mathrm{B} B \gg k_\mathrm{B} T$ we have $\zeta = 1$, i.e., the tunnel amplitudes are fully polarized for spin-up carriers, see equation (S-17), whereas for $g\mu_\mathrm{B} B \ll k_\mathrm{B} T$ we have $\zeta = 0$. This spin-dependence is physically motivated by the fact that our quantum dot is contacted by CNT leads as shown in Fig. S-3b and explained in the main article [cf. Fig. 2b]. When increasing the field by a few Tesla the strong spin-polarization in the tunnel amplitudes suppresses the passage of spin-down electrons through the dot and hence the $T_0^0$ and $S_0$ lines (together with their vibrational sidebands) are suppressed since the ground state in the $N = 1$ sector is $D_-^\uparrow$. What remains are the $T_0^+$ excitation and its vibrational sidebands.

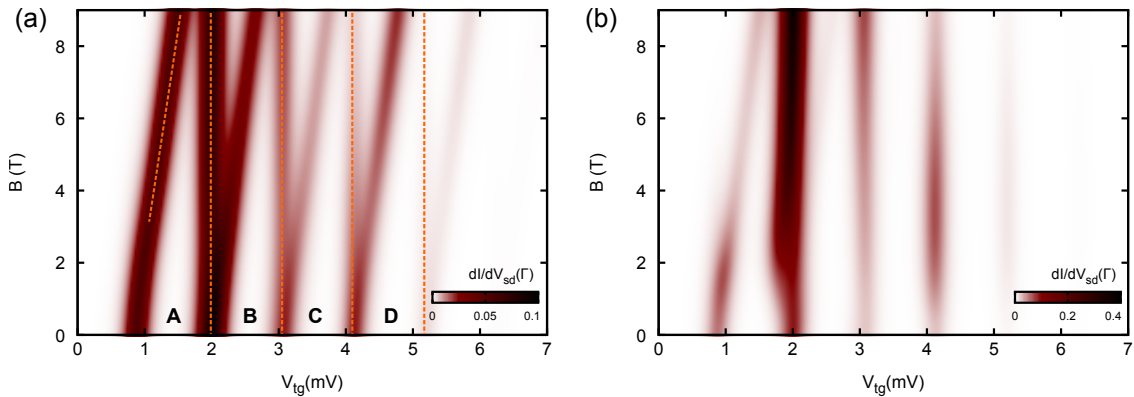

FIG. S-10. Calculated magnetic field evolution for the plotted differential conductance of Fig. 2d of the main article: (a) Spin-independent tunnel amplitudes ($\zeta = 0$ for all $B$). (b) Spin-dependent tunnel amplitudes according to equation (S-59). Further parameters [same as in Fig. 2g of the main article]: $\Delta = 0.8$ meV, $J = 1.5$ meV, $\lambda_\varepsilon = 0.28$, $\lambda_\Delta = 0.32$, $\lambda_J = -0.22$, $\hbar\omega = 0.85$ meV, $\Delta_{SO} = 0.1$ meV, $\kappa = 0.3$, $\gamma = -0.5$ and $T = 0.7$ K.

However, the ground singlet $S_-$ is also suppressed when assuming this spin-dependence, since it requires an additional spin-down electron to fill the bonding orbital ($\tau = -$), cf. equation (S-9). The presence of the ground singlet $S_-$ line, but not its vibrational sidebands is the key feature of the experiment. It is at this point that the spin-orbit coupling does have a decisive effect: *when even a small spin-orbit coupling is included, the singlet $S_-$ reappears by borrowing intensity from the triplet $T_0$ (see also Fig. 4), but without reinstating the unobserved $S_0$ and the Zeeman split-off states of $T_0$ and their vibrational sidebands.* The SO coupling mostly mixes the ground singlet $S_-$ and the triplet $T_0^1$ states [cf. Sec. II A 3]. In Fig. S-10b [same data as in Fig. 3c of the main article] we show the magnetic field evolution of the calculated differential conductance along the line $V_{sd} = V_{tg} + 5$ mV when we include the finite $\zeta$ of equation (S-59): indeed the suppression of the $T_0^0$ and $S_0$ lines and their sidebands is maintained while only the ground $S_-$ has a clear intensity. This produces the observed intensity pattern of Fig. 3a of the main article. It is at this last point that the Anderson-Holstein model fails, see Sec. II C 3.

*From this we infer that in our device an interplay of spin-filtering of the suspended CNT parts, functioning as tunnel junctions, and weak spin-orbit interaction in the CNT quantum dot may be responsible for the suppression of Zeeman splitting.* As emphasized above it is not responsible for the essential effect: The missing vibrational side bands while having a clear singlet ground state $S_-$ line.

### 3. Influence of the various parameters and the problem of Anderson-Holstein coupling

Having made the detailed comparison with the experiment we now outline the different influence the various parameters on the differential conductance. We illustrate this for the case of an Anderson-Holstein type model ($\lambda_\Delta = \lambda_J = 0$) and expose the problem it has in explaining the data.

In Figures S-11a - S-11c we plot the results obtained for our model, keeping all parameters as in the main article, except for $\gamma$ which we vary and making the Anderson-Holstein approximation by setting $\lambda_\Delta = \lambda_J = 0$. Also, to obtain a similar number of visible vibrational sidebands we have to choose a larger value of $\lambda_\varepsilon = 1$ in this case. To find agreement with the experiment the lines with positive slope should first of all be suppressed: these are vibrational sidebands relating to the $|D_-, \nu\rangle \leftrightarrow |S_-, 0\rangle$ where $\nu = 1, 2, 3, ..$ quanta are excited for $N = 1$. This can be done by increasing the junction asymmetry $\gamma$ a lot. The best result obtainable this way, Fig. S-11c should now be compared with Fig. 2g of the main article: as one adjusts $\gamma$ to suppress the vibrational sidebands relating to the $S_-$ (both with positive and negative slope) the ground state $\nu = 0$ transition also becomes suppressed, resulting in a severe current blockade at low bias $V_{sd} \lesssim \hbar\omega$. In contrast, in Fig. 2g these lines are not present due to the state-dependent vibrational coupling

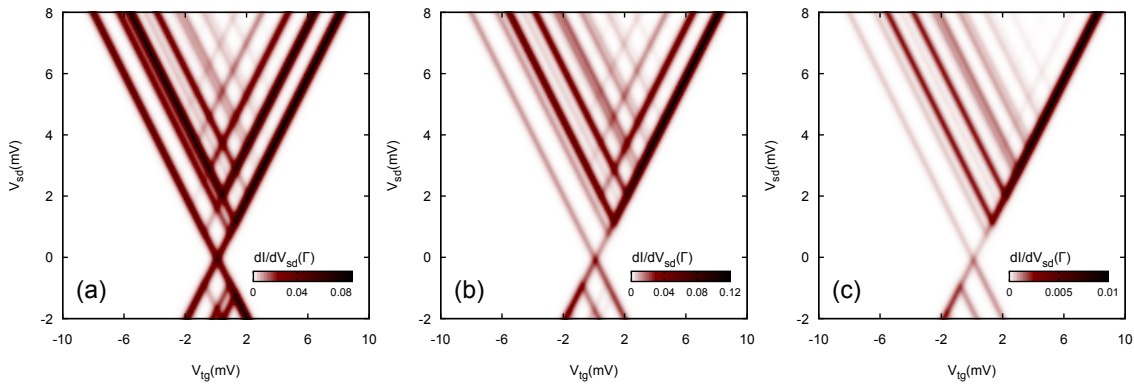

FIG. S-11. Differential conductance at $B = 0$ for the Anderson-Holstein model ($\lambda_\Delta = \lambda_J = 0$) with $\lambda_\varepsilon = 1$. The other parameters are the same as for the state-dependent vibrational coupling model, i.e., $\Delta = 0.8$ meV, $J = 1.5$ meV, $\hbar\omega = 0.85$ meV, $\kappa = 0.3$, $\zeta = 0.0$ and $T = 0.7$ K. The panels show results for increasing junction asymmetry $\gamma$: (a) $\gamma = 0$, (b) $\gamma = -0.5$ and (c) $\gamma = -0.9$. For completeness we have also included here the SO coupling $\Delta_{SO} = 0.1$ meV as we also did in our model with state-dependent coupling ($\lambda_\Delta, \lambda_J \neq 0$), although this has little effect at $B = 0$.

$\lambda_\Delta$ and $\lambda_J$ which does not block the transport at low voltage. This allows for a much smaller asymmetry, leaving the ground-state transition fully visible.

This problem becomes more pressing when we now consider the magnetic field evolution: while $S_-$ and its sidebands with negative $V_{tg}$-slope are still present they are quasi-degenerate with the triplet excitations at $B = 0$ in panel (c). Turning on the field $B$ they will however split off from the triplet by the Zeeman effect. Also in this case one must assume a spin-dependent tunneling $\zeta$ [equation (S-59)] to find agreement with the experimental data of Fig. 3a where the Zeeman-split off states are not observed. As before, this also suppresses the $S_-$ transition, both the ground one ($\nu = 0$) and all vibrational sidebands ($\nu > 0$). However, when turning on the SO coupling again to remedy this, all $S_-$ transition are restored, i.e., including the vibrational side bands (this happens again by borrowing intensity from the $T_0$ triplet). This disagrees with the key experimental observation that only the triplet transitions show vibrational side bands.

The above discussion underlines the importance of the experimental advance reported in the main article: by being able to switch "on" and "off" the vibrational coupling as well as performing magnetic field transport spectroscopy of the quantum dot states, we are able to identify electronic states with different coupling to the vibration.

Next, by varying orbital asymmetry in the tunneling $\kappa$ one changes the magnitude of the $S_-$ transitions relative to the $T_0$ and $S_0$ transitions: when enhancing $\kappa$, the tunneling in the antibonding ($|-\rangle$) orbital is enhanced: since $S_-$ has two electrons in that orbital, whereas $T_0$ and $S_0$ have only one, this enhances the former relative to the latter. As above, $\kappa$ cannot be adjusted to find agreement with the experiment: as in the discussion of the $\zeta$ dependence, it cannot suppress the vibrational side bands of $S_-$ relative to the $S_-$ transition: this requires state-dependent coupling.

*Thus the key problem in trying to use an Anderson-Holstein model to explain state-dependent vibration coupling lies in its basic assumption that all electronic states with the same charge couple equally to the vibration.* It fails because it either does *not at all* show the $D_- \to S_-$ excitation (i.e., neither the ground ($\nu = 0$) or any vibrational sidebands $\nu \neq 0$) or it *does* show it together with *all* its vibrational sidebands. Which of the two is the case depends on parameters, but an apparent state-dependent vibrational coupling seems impossible to achieve. The sidebands for the singlet $S_-$, relative to the ground transition, are intense as those of the triplet $T_0$ relative to its ground transition. Neither junction ($\gamma$), orbital ($\kappa$) nor spin-asymmetry ($\zeta$) can get around this fact.

---


\end{document}